# Few-Shot Specific Emitter Identification via Deep Metric Ensemble Learning

Yu Wang, *Graduate Student Member*, *IEEE*, Guan Gui, *Senior Member*, *IEEE*, Yun Lin, *Senior Member*, *IEEE*, Hsiao-Chun Wu, *Fellow*, *IEEE*, Chau Yuen, *Fellow*, *IEEE*, and Fumiyuki Adachi, *Life Fellow*, *IEEE*

**Abstract**

Specific emitter identification (SEI) is a highly potential technology for physical layer authentication that is one of the most critical supplement for the upper-layer authentication. SEI is based on radio frequency (RF) features from circuit difference, rather than cryptography. These features are inherent characteristic of hardware circuits, which difficult to counterfeit. Recently, various deep learning (DL)-based conventional SEI methods have been proposed, and achieved advanced performances. However, these methods are proposed for close-set scenarios with massive RF signal samples for training, and they generally have poor performance under the condition of limited training samples. Thus, we focus on few-shot SEI (FS-SEI) for aircraft identification via automatic dependent surveillance-broadcast (ADS-B) signals, and a novel FS-SEI method is proposed, based on deep metric ensemble learning (DMEL). Specifically, the proposed method consists of feature embedding and classification. The former is based on metric learning with complex-valued convolutional neural network (CVCNN) for extracting discriminative features with compact intra-category distance and separable inter-category distance, while

Y. Wang and G. Gui are with the College of Telecommunications and Information Engineering, Nanjing University of Posts and Telecommunications, Nanjing 210003, China (E-mails: {1018010407@njupt.edu.cn, guiguan@ @njupt.edu.cn).

Y. Lin is with the College of Information and Communication Engineering, Harbin Engineering University, Harbin 150000, China (E-mail: linyun@hrbeu.edu.cn).

H.-C. Wu is with the School of Electrical Engineering and Computer Science, Louisiana State University, Baton Rouge, LA 70803, USA (E-mail: wu@ece.lsu.edu).

C. Yuen is with the Engineering Product Development (EPD) Pillar, Singapore University of Technology and Design, Singapore 487372, Singapore (E-mail: yuenchau@sutd.edu.sg).

F. Adachi is the Research Organization of Electrical Communication, Tohoku University, Sendai 980-8577 Japan (E-mail: adachi@ecei.tohoku.ac.jp).

the latter is realized by an ensemble classifier. Simulation results show that if the number of samples per category is more than 5, the average accuracy of our proposed method is higher than 98%. Moreover, feature visualization demonstrates the advantages of our proposed method in both discriminability and generalization. The codes of this paper can be downloaded from GitHub[1].

## I. Introduction

Wireless communication is rapidly developed for realizing the great vision of the internet of everything. Meanwhile, there are various unprecedented challenges, especially the security, because of the open nature of wireless channel [1]–[3]. Authentication is one the most critical techniques for the secure wireless communications, and the existing authentication methods rely on the cryptography-based authentication mechanisms above physical layer [4], [5]. However, since the inherent defects of these cryptography-based authentication mechanisms, such as vulnerable to spoofing, forgery, replay, eavesdropping and reflection attacks [6], [7], the requirement of additional communication overhead and high complexity [8], they may not be applicable for emerging wireless communication systems.

Therefore, physical layer authentication, as an effective supplement for the upper-layer authentication, is widely concerned [9]. In this paper, we mainly focus on a passive physical layer authentication technology, named as specific emitter identification (SEI), which is a greatly potential technology in both military and civilian scenarios. SEI relies on radio frequency (RF) features, usually parasitic in transmitted signals and originate from hardware differences from circuits [10], for the identification of different transmitters. Due to that the hardware difference is an inherent attribute in the process of circuit design or circuit manufacturing, RF features are tamper-resistant and difficult to counterfeit, which is also the basis of SEI.

The typical RF feature-based SEI methods generally consist of pre-processing, feature extraction and identification [11]. The pre-processing usually contains various operations, including filtering, power normalization, synchronization, target signal interception and so on [12], while the feature extraction and the identification are the key steps of SEI. Before the development of deep learning (DL), classical SEI methods mainly design features through statistical methods, and then machine learning-based classifiers are applied for the identification. However, there are many disadvantages of the RF features designed by artificial knowledge and experience, such as the weak performances, the poor environmental adaptability and the lack of capabilities to deal with increasingly complex and increasing wireless transmitters.

[1]https://github.com/BeechburgPieStar/Few-Shot-Specific-Emitter-Identification-via-Deep-Metric-Ensemble-Learning

In recent years, DL has been proved to have the powerful and effective data analysis capability [13], it has been widely applied into wireless communication technologies [14]–[16], [18], including SEI [19]–[31]. These DL-based SEI methods relied on massive historical RF signal samples and deep neural networks for extracting more robust and effective RF features, which have been demonstrated to have better performances than artificial feature-based methods.

However, most of the DL-based SEI methods focused on the close-set SEI problem [19]–[31], which will be specifically introduced in Section II. In other words, the transmitter categories are the same in the stages of offline training and online deployment, but it is invalid in actual applications. Because it is scarcely possible to collect signals samples from all categories of transmitters for training. Although this problem can be solved by finetuning, it is constraint by few training signal samples of new categories and limited retraining time in the process of online deployment. Hence, we focus on the few-shot SEI (FS-SEI) for aircraft identification task by automatic dependent surveillance-broadcast (ADS-B) signals, which are applied for aircraft monitoring [32]. There are massive ADS-B signal samples from multiple aircrafts as auxiliary dataset for the offline training, and few ADS-B signal samples from new aircrafts as few-shot training samples in the stage of the online deployment. In this paper, a deep metric ensemble learning (DMEL)-based SEI method is proposed, consisting of feature embedding based on metric learning and complex-valued convolutional neural network (CVCNN) and ensemble learning-based classifier.

Specifically, the former aims to extracts discriminative features from ADS-B signal samples by combining Softmax loss and two metric-based contrastive losses (including triplet loss and center loss). Different from the separable features extracted by the conventional DL methods with Softmax loss, discriminative features is more generalized, which is not only applicable to the ADS-B signal samples from seen categories, but also suitable to the ADS-B signal samples from unseen categories or categories that are rarely seen. In addition, based on the discriminative features, the identification of aircrafts in the latter is easy, but we also introduce ensemble learning to combine multiple machine learning classifiers for further enhancing the identification performance. The main contributions of this paper are listed as follows.

- We introduce the paradigm of "constructing feature embedding offline and establishing classifier online" for FS-SEI. In the offline training phase, feature embedding is constructed by CVCNN and massive historical ADS-B signal samples, because the feature embedding usually contains large numbers of hyper-parameters, which requires to be trained by massive samples. In the online deployment phase, a simple classifier with few hyper-parameters or even no hyper-parameters can be established by limited ADS-B samples from new categories of aircrafts.
- We propose a metric learning based feature embedding scheme for extracting generalized and discriminative features. Here, we combine two contrastive losses and Softmax loss to ensuring compact intra-category distance and separable inter-category distance in the feature space.
- We use a simple machine learning classifier followed by the above good feature embedding to realize identification, and ensemble learning is introduced for further improving identification performance.

## II. Related Works

In this section, conventional artificial features and DL-based SEI methods and four few-shot learning (FSL) algorithms with their applications are introduced.

### *A. Artificial RF Feature-Based SEI Methods*

Artificial RF feature-based SEI methods are usually based on artificial RF features and machine learning-based classifiers, and there are various RF features for SEI, which can be mainly summarized as instantaneous features, modulation domain features and transform domain features, which are introduced as follows.

The first kind of RF features is the instantaneous statistics of RF signal, and it usually contains the mean, variance, skewness, and kurtosis of amplitude, phase and frequency about signals or signals reconstructed by variational mode decomposition (VMD) [33]. Next, modulation domain features are nonlinear characteristics introduced in the modulation process, including in-phase

TABLE I

DL MODELS AND THE NUMBER OF RF SIGNAL SAMPLES IN CONVENTIONAL DL-BASED SEI METHODS.

| **Conventional methods** | **Model** | **Transmitter type** | **Data format** | **Number of samples** |
|---|---|---|---|---|
| K. Merchant *et al.* [19] | CNN | 7 Zigbee devices | Raw IQ | 119,000 (90% for training and validation) |
| J. Yu *et al.* [20] | MSCNN | 54 Zigbee devices | Raw IQ | 62,860 (80% for training and validation) |
| X. Wang *et al.* [21] | LSTM | 4 RF devices | Raw IQ | 12,000 (training)/250 (transferring) |
| B. He and F. Wang [22] | LSTM | 5 Emitters | Signal components | 500 samples per device for training |
| T. Jian *et al.* [23] | ResNet | 500 WIFI devices and 50 ADS-B transmitters | Raw IQ | 218 transmissions per device |
| Y. Wang *et al.* [24] | CVCNN | 7 Power amplitudes | Raw IQ | 262,000 samples (70% for training) |
| L. Ding *et al.* [25] | CNN | 5 USRPs | Bispectrum | 3,000 samples (50% for training) |
| L. Ding *et al.* [26] | CNN | 20 LoRa devices | Bispectrum | 1,000 packets per device for training |
| C. Xie *et al.* [27] | CNN | 6 USRPs | Bispectrum | 10,000 unlabeled samples per category and 200∼1200 labeled samples per category |
| J. Gong *et al.* [28] | InfoGAN | 5 routers | Gray histogram | 50,000 samples |
| L. Peng *et al.* [29] | CNN | 54 Zigbee devices | DCTF | 8,262 samples |
| P. Yin *et al.* [30] | MCCNN | 6 LTE mobile phones | Multi-part DCTF | 2,540 samples (80% for training) |
| Y. Peng *et al.* [31] | CNN | 7 Power amplitudes | HCTF | 310∼410 samples per device |

and quadrature (IQ) imbalance [34], modulation shape [35], constellation error [36] and so on. Finally, transform domain features are the features based on various transforms, including wavelet transform [37], Hilbert-Huang transform [38], short-time Fourier transform [39] and so on.

### *B. DL-Based SEI Methods*

DL-based SEI methods apply neural networks, such as convolutional neural network (CNN), recurrent neural network (RNN), generative adversarial network (GAN), for joint feature extraction and classification. Here, the DL-based SEI methods can be mainly summarized into three categories by sample format, i.e., time domain signal, spectrogram and constellation.

*1) Time domain signal-based SEI methods:* Time domain signals refer to raw IQ signals or signal components decomposed by empirical mode decomposition (EMD), variational mode decomposition (VMD) and so on. K. Merchant *et al.* [19] firstly introduced a multi-layer CNN for SEI based on seven Zigbee devices and their raw IQ signal samples, which demonstrated that CNN can effectively distinguish different transmitters, even for different devices with the same type and the same model. Similarly, J. Yu *et al*. [20] also proposed a multi-sampling CNN (MSCNN) for SEI based on raw IQ signal samples, using multiple downsampling transformations for joint multi-scale feature extraction and classification. It is worth noting that they applied 54 Zigbee devices for simulations, and their proposed MSCNN has achieved 97% accuracy at high SNR, which illustrated the effectiveness of DL in SEI.

Different from the above CNN-based SEI methods, X. Wang *et al*. [21] proposed a LSTM and raw IQ sample-based SEI method, but they focused on the problem that the identification performance of SEI declines with time, and they introduced transfer learning for solving this problem. B. He and F. Wang [22] also proposed a SEI method based on LSTM, but their method applied signal components decomposed by EMD, intrinsic time-scale decomposition (ITD), or VMD as training or test samples. More importantly, they explored the improvement of identification performance by multiple receivers.

Although the above DL and time domain signal-based SEI methods have achieved advanced identification performances, their neural network structures are designed by artificial experience, which usually bring into inefficient identification, becasue of redundant neurons or invalid structures. So, there are some DL-based SEI methods, whose DL structures are designed by data-driven methods, such as sparse structure selection [23], [24], for better performance or faster inference speed. T. Jian *et al*. [23] and Y. Wang *et al*. [24] designed the neural networks by constructing the original model and then removing redundant neurons for sparse structures with regularization. The difference is that the former applied the real-valued CNN and the optimization method is alternating direction method of multipliers (ADMM), while the latter focused on the complex-valued CNN (CVCNN) and it is optimized by proximal gradient descent. Simulation results demonstrated that there is the redundance in model parameters of the previous proposed DL-based SEI methods.

*2) Spectrogram-based SEI methods:* In these spectrogram-based SEI methods, bispectrum analysis and Fourier transform are the common solutions to transform signals into spectrograms. L. Ding *et al.* [25] firstly applied compressed bispectrums by supervised dimensionality reduction method as samples, and they used a simple four-layer CNN as identification method, which has achieved nearly 100% accuracy at high SNR. G. Shen *et al.* [26] proposed a short-time Fourier transform-based spectrogram and CNN-based SEI method, which has been demonstrated to have the better identification performance than IQ-based methods and fast Fourier transform-based methods. In addition, the influence of carrier frequency offset (CFO) on the identification performance is also revealed, and the estimated CFO is introduced to modify the identification result, so as to avoid the performance degradation.

Different from the previous supervised methods, C. Xie *et al.* [27] and J. Gong *et al.* [28] proposed the semi-supervised and unsupervised SEI methods, respectively. Specifically, the former used the bispectrum as the sample and virtual adversarial training-based CNN for realizing semi-supervised SEI; The latter is based on information maximized GAN (InfoGAN) with the priori information of the wireless channels, and the input for this method is the gray histogram constructed by the bispectrum, which is to enhance the discriminability between different transmitters.

*3) Constellation-based SEI methods:* L. Peng *et al.* [29] designed a novel differential constellation trace figure (DCTF) as the input of CNN, and their simulation results demonstrated that these signal representation methods are useful for SEI. Based on [29], P. Yin *et al.* [30] proposed a multi-channel CNN (MCCNN)-based SEI method for wireless terminal authentication, which used multi-part DCTFs as training and test samples. In detail, a set of multi-part DCTFs is composed of three DCTFs from the transient-on, modulation and transient-off parts, respectively. In the training and test phase, these three DCTFs are fed into three independent CNNs, and then the features extracted from these CNNs are concatenated for classification. This simulation is based on six long term evolution (LTE) mobile phones and their simulation results indicated that multi-part DCTF is better than single-part DCTF.

Except DCTF, Y. Peng *et al.* [31] introduced a novel heat constellation trace figure (HCTF) for SEI. Specifically, the principle of HCTF is that the distribution density of different regions is calculated on the basis of the common constellation trace figure, and the regions with different densities are given different colors. The simulation results demonstrated that HCTF with Inception v3 has the advanced and robust performance under various channel conditions.

The reason why the above DL-based SEI schemes can achieve advanced performance is not only their excellent deep learning algorithms, but also the large number of RF signal samples. J. Gong *et al.* [28] revealed the relationship between the number of training samples and the identification performance in their paper, i.e., the DL-based SEI method need sufficient samples, otherwise its performance is far worse than the traditional feature-based methods. Specifically, if there are 500 samples per device for training, the LSTM and ITD-based SEI method can have 96.70% accuracy, but the accuracy will decline to 65.50%, if there are only 10 samples per device for training. The detailed information are given in TABLE I. Clearly, the number of the training samples per category are usually hundreds, thousands, or even tens of thousands. If there are not enough samples in the above methods, they may hardly achieve these excellent results.

### *C. Typical FSL Methods and Their Applications*

There are four typical FSL methods, including data augmentation, generative model, metric model and meta model [40], which are specifically introduced below. In addition, the applications are only related to the signal identification.

*1) Data augmentation:* Data augmentation aims to expand the few-shot training dataset manually by making limited samples produce more equivalent samples, and it is an effective method to overcome the lack of training samples, but no matter what kind of expansion schemes, it will bring the problem of sample noise. H. Zhou *et al*. [41] proposed a data union augmentation method for modulation signal classification. In detail, authors applied GAN to generate samples, and then selected the generated samples with high similarity with the original samples for reducing the impact of sample noise. In addition, the samples are further expanded by spatiotemporal flip. Simulation results demonstrated that this method can improve classification performance under the condition of limited samples.

It is noted that most of data augmentation methods are only based on few-shot training datasets from their tasks to expand samples, but other three schemes will introduce auxiliary datasets from various auxiliary tasks. Due to the additional information brought by the auxiliary datasets, the performance of the other three schemes is usually better than that of data augmentation, but data augmentation can be a trick to be added into other FSL methods.

*2) Generative model:* Generative methods applies auto-encoder or GAN to extract robust features, which can represent and recover the corresponding samples. For instance, Y. Wang *et al*. [42] used a feature Wasserstein GAN for RF-based human activity recognition under the few shot condition, which can recognize new activities with high accuracy using few samples.

*3) Metric model:* Metric methods are based on the ideal of “learn to compare” or “learn to measure”, which aims to compare or measure the similarity between different samples. One of the most famous metric methods is the Siamese network (SiameseNet) [43]. SiameseNet is a coupling framework based on two weight-sharing sub neural networks. Paired samples are respectively fed into sub neural networks for their corresponding features, and then their similarly can be compared by calculating the distance between features, such as Euclidean distance. The detailed structure of SiameseNet is shown as follows.

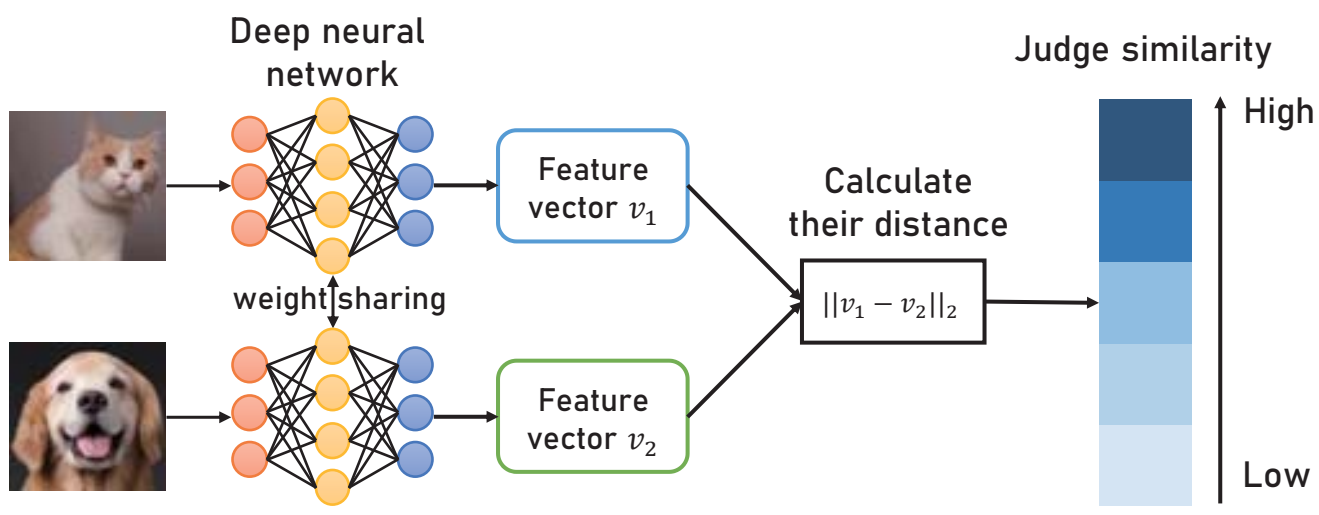

Fig. 1. The structure of SiameseNet based on Euclidean distance.

SiameseNet can effectively distinguish the sample from one categories from other categories. Y. Wu *et al.* [44] and G. Sun [45] introduced it into SEI for the tasks of open-set SEI and FS-SEI, respectively. In addition, Z. Zhang *et al.* [46] and P. Man *et al.* [47] applied the relation network, which is an improved SiameseNet and uses a fully connected neural network to judge the degree of similarity, for signal modulation recognition and zero-shot SEI.

*4) Meta model:* The basic ideal of meta methods is "learn to learn". The essence of meta learning is an optimization method to increase the generalization of neural networks in the multi-tasks, and its purpose is to adapt new tasks on the basis of acquiring existing knowledge. N. Yang *et al.* [48] proposed a model-agnostic meta learning (MAML)-based SEI method, and their simulation on the identification of different types of transmitters demonstrated its ultra-high performance under the condition of limited samples. Y. Dong *et al.* [49] also applied MAML with complex-valued attention for modulation recognition, which also achieved amazing recognition performance.

Compared with metric methods, meta methods, especially MAML, has the the disadvantages of high training complexity and unstable gradient [50], because of their complex training process. Due to the training processing of the metric methods is simple, their implementation is not much different from DL, and these methods just need to modify loss function. Thus, we focus on the metric methods in this paper.

# III. System Model, Dataset Generation and Problem Description

## A. System Model

In this paper, SEI technology is applied to identify aircrafts based on ADS-B signals. The SEI system for different aircraft identity authentication is shown in Fig. 2, and there are three steps: data collection, model training and model deployment. In general, data collection is conducted in the previous airspaces, and SEI models are trained on the historical ADS-B signal samples from these airspaces, but model deployment is usually conducted in a new airspace.

Considering that there are new aircrafts in the new airspace that have never appeared in the previous airspaces, it is difficult to directly deploy the SEI model based on historical ADS-B signal samples from the previous airspace in the new airspace. In detail, there are at least two problems for the quick deployment: (1) It is scarcely possible to collect enough supervised ADS-B signal samples from new aircrafts for SEI model updating in a short time; (2) Even if ADS-B signal samples are sufficient, it is difficult to quickly update the SEI model for identifying new aircrafts. Therefore, under the condition of limited ADS signal samples, FSL is introduced for rapid update and real-time deployment of SEI.

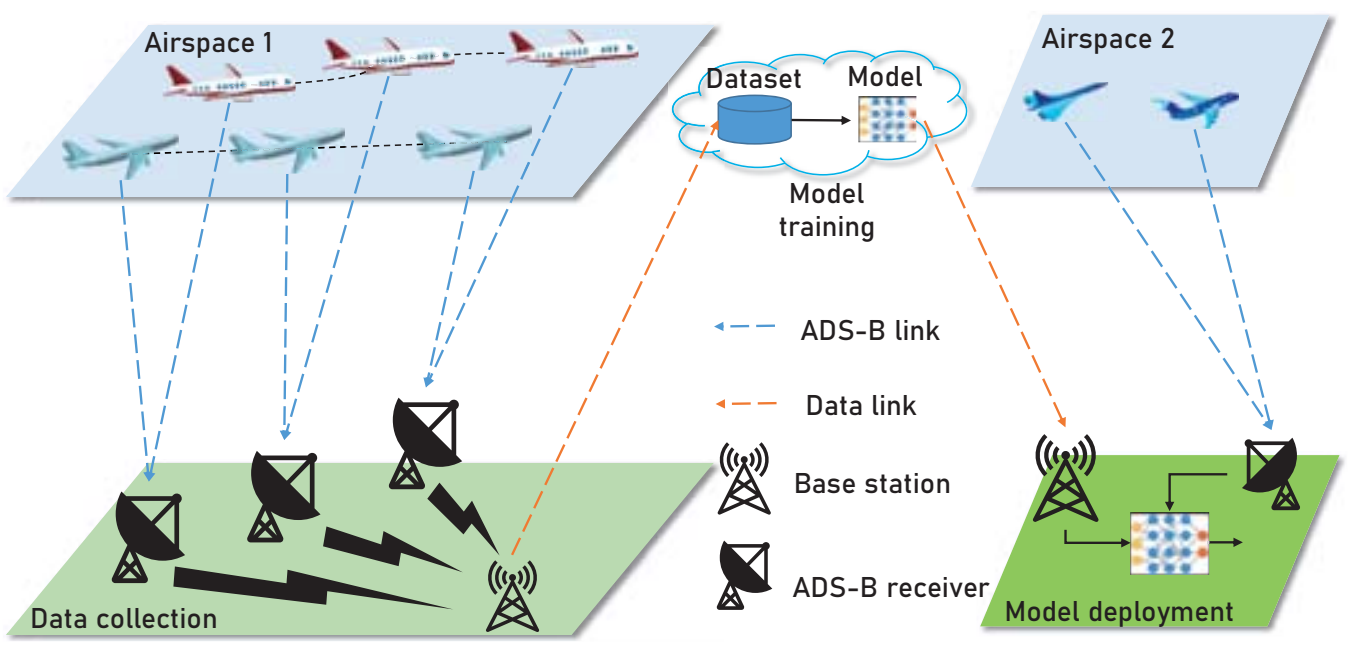

Fig. 2. The system structure of SEI-based aircraft identity authentication: data collection, model training and model deployment.

## B. ADS-B Signal Dataset

The received ADS-B signal model can be written as follows.

$$x(t) = h(t) * s(t) + n(t), \tag{1}$$

where $x(t)$ is the received signal; $h(t)$ and $n(t)$ represent wireless channel and noise, respectively; $s(t)$ is the transmitted ADS-B signal, which consists of the preamble and the pulse position modulation-based data block. There are four pulses with fixed positions and $0.5 \pm 0.05$ $\mu$s duration per pulse in the preamble, and the duration of the data block is $112$ $\mu$s.

The ADS-B signal collector consists of Signal Hound SM200B and 1090 MHz omnidirectional antenna. In addition, high performance computing terminal is applied to detect ADS-B signal from original signal data, and construct the ADS-B signal dataset. Specifically, the data block contains aircraft identification code, position, altitude, speed and so on. So, sample category (i.e., aircraft identification code) can be obtained by the demodulation of the data block. Moreover, the pre-processed ADS-B signals are as IQ samples. The more detailed information of the ADS-B signal dataset has been open-source, which can refer to [51].

### *C. Problem Description*

Here, $\mathbf{x}$ represents the input sample, which is a ADS-B signal sample from one aircraft with IQ component format; $\mathrm{y}$ represents the real category of the corresponding aircraft; $\mathbf{D} = \{\mathbf{x}_n, \mathrm{y}_n\}_{n=1}^N$ denotes the dataset containing ADS-B signal samples and their corresponding categories. In addition, $\mathbf{x}_n$ comes from a specific domain $\mathcal{D} = \{\mathcal{X}, P_{\mathcal{X}}\}$, where $\mathcal{X}$ ($\mathbf{x}_n \in \mathcal{X}$) is the sample space and $P_{\mathcal{X}}$ is the marginal probability distribution of the sample space. Moreover, we use $\mathcal{Y}$ ($\mathrm{y}_n \in \mathcal{Y}$) to represent the category space, and $P_{\mathcal{X}\times\mathcal{Y}}$ to represent the joint probability distribution of the sample space and the category space.

*1) SEI problem:* SEI problem can be defined as a maximum-a-posteriori (MAP) criterion-based multi-class pattern recognition problem, which can be written as

$$\hat{\mathrm{y}} = \arg\max_{\mathrm{y}\in\mathcal{Y}} f_{\mathrm{SEI}}(\mathrm{y}|\mathbf{x}; \mathbf{W}), \tag{2}$$

where $\hat{\mathrm{y}} = f_{\mathrm{SEI}}(\mathbf{x}; \mathbf{W})$ is the predicted category, and $f_{\mathrm{SEI}}$ and $\mathbf{W}$ are the mapping function and the set of hyper-parameters for SEI, respectively. When the mapping function is fixed, the goal of SEI is to find a set of suitable hyper-parameters $\mathbf{W}$ such that it can realize the mapping from the sample space to the category space, i.e., $f_{\mathrm{SEI}}(\mathbf{W}) : \mathcal{X} \rightarrow \mathcal{Y}$. $\mathbf{W}$ belongs to the hyper-parameter space $\mathcal{W}$, and it can minimize the expected error $\varepsilon_{ex}$, i.e.,

$$\min_{\mathbf{W}\in\mathcal{W}} \varepsilon_{ex} = \min_{\mathbf{W}\in\mathcal{W}} \mathbb{E}_{(\mathbf{x},\mathrm{y})\sim P_{\mathcal{X}\times\mathcal{Y}}} \mathcal{L}(\hat{\mathrm{y}}, \mathrm{y}), \tag{3}$$

where $\mathcal{L}(\cdot)$ is to measure the difference between the predicted category and the ground-truth category. However, $P_{\mathcal{X}\times\mathcal{Y}}$ is usually unknown, and thus DL algorithms generally minimize the empirical error $\varepsilon_{em}$ as the replacement of $\varepsilon_{ex}$, which can be written as

$$\min_{\mathbf{W}\in\mathcal{W}} \varepsilon_{em} = \min_{\mathbf{W}\in\mathcal{W}} \mathbb{E}_{(\mathbf{x},\mathrm{y})\sim\mathbf{D}}\mathcal{L}(\hat{\mathrm{y}},\mathrm{y}). \tag{4}$$

Moreover, the generalization error between $\varepsilon_{ex}$ and $\varepsilon_{em}$ is $\varepsilon_{ge} = |\varepsilon_{em} - \varepsilon_{ex}|$. If the hyper-parameter space $\mathcal{W}$ to be searched is too huge, $\varepsilon_{ge}$ will be large, and there is a high probability of overfitting. In order to minimize $\varepsilon_{ge}$, many constraints will be imposed on $\mathbf{W}$ to narrow the hyper-parameter space $\mathcal{W}$, and thus the above problem can be written as

$$\begin{aligned} &\min_{\mathbf{W}\in\mathcal{W}} \varepsilon_{em}, \\ &\text{s.t. } f_{\mathrm{SEI}}(\mathbf{x}_n;\mathbf{W}) = \mathrm{y}_n, \forall(\mathbf{x}_n,\mathrm{y}_n)\in\mathbf{D}. \end{aligned} \tag{5}$$

It is obvious that more supervised samples in $\mathbf{D}$ represent more constraints imposed on $\mathbf{W}$, and the hyper-parameter space will be smaller, which may bring in an excellent generalization performance. Thus, a generalized DL model can generally be achieved by the simple supervised learning algorithm under the condition of sufficient supervised samples, but it is difficult to train the similar model in case of insufficient supervised samples, and this is the problem to be solved by FSL.

*2) FS-SEI problem:* FS-SEI problem is how to train an excellent SEI method with few ADS-B signal training samples. This problem can be described by a set of few-shot ADS-B signal datasets, consisting of training dataset and test dataset, i.e., $\mathbf{D}_{fs} = \{\mathbf{D}_{tr},\mathbf{D}_{te}\}$. $\mathbf{D}_{tr} = \{(\mathbf{x}_i,\mathrm{y}_i)\}_i^{N_{tr}}$ contains $C$ categories with $K$ samples per category, where $N_{tr} = C\times K$ and $K$ is usually small (for instance, $K=1$ or 5), while there are sufficient samples of the same category as $\mathbf{D}_{tr}$ in $\mathbf{D}_{te} = \{(\mathbf{x}_j,\mathrm{y}_j)\}_j^{N_{te}}$, where $N_{te} \gg N_{tr}$. In addition, it is noted that $\mathbf{x}_i,\mathbf{x}_j \in \mathcal{X}_{fs}\subset\mathcal{X}$ and $\mathrm{y}_i\in\mathcal{Y}_{fs}\subset\mathcal{Y}$, where $\mathcal{X}_{fs}$ and $\mathcal{Y}_{fs}$ are the few-shot sample space and the few-shot category space.

However, according to formula (5), it is almost impossible to construct an effective $f_{\mathrm{SEI}}(\mathbf{W})$ only based on $\mathbf{D}_{te}$ with extremely limited constraints. Thus, an auxiliary dataset $\mathbf{D}_{au}$ is generally introduced for narrowing the hyper-parameter space $\mathcal{W}$ and achieving generalized model. In general, $\mathbf{D}_{au} = \{(\mathbf{x}_a,\mathrm{y}_a)\}_a^{N_{au}}$ is a supervised dataset containing sufficient historical ADS-B signal samples with their corresponding categories, i.e., $N_{au} \gg N_{tr}$ and $|\mathcal{Y}_{au}| \gg |\mathcal{Y}_{fs}|$. Here, $\mathbf{x}_a\in\mathcal{X}_{au}\subset\mathcal{X}$ and $\mathrm{y}_a\in\mathcal{Y}_{au}\subset\mathcal{Y}$, where $\mathcal{X}_{au}$ and $\mathcal{Y}_{au}$ are the auxiliary sample space and the auxiliary category space.

Moreover, it is noted that the categories in $\mathbf{D}_{fs}$ do not appear in $\mathbf{D}_{au}$, i. e., $\mathcal{Y}_{fs} \cap \mathcal{Y}_{au} = \emptyset$, but the few-shot ADS-B signal sample $\mathbf{x}_i$ and auxiliary ADS-B signal sample $\mathbf{x}_a$ come from the same domain, that is, $\mathcal{D}_{fs} = \mathcal{D}_{au}$, $\mathcal{X}_{fs} = \mathcal{X}_{au}$ and $P_{\mathcal{X}_{fs}} = P_{\mathcal{X}_{au}}$, where $\mathcal{D}_{fs} = \{\mathcal{X}_{fs}, P_{\mathcal{X}_{fs}}\}$ and $\mathcal{D}_{au} = \{\mathcal{X}_{au}, P_{\mathcal{X}_{au}}\}$ are the few-shot domain and the auxiliary domain, respectively, and $P_{\mathcal{X}_{fs}}$ and $P_{\mathcal{X}_{au}}$ are the marginal probability distributions of the few-shot sample space and the auxiliary sample space, respectively.

In general, the idea of FS-SEI is to construct $f_{\text{SEI}}(\mathbf{W})$ with the limited supervised knowledge in $\mathbf{D}_{tr}$ and the auxiliary knowledge in the $\mathbf{D}_{fs}$-irrelevant dataset $\mathbf{D}_{au}$. The assumption for FS-SEI or other FS problems is reasonable and feasible, since that $\mathbf{D}_{au}$ is easily available.

## IV. The Proposed FS-SEI Method Based on DMEL

### *A. The Framework of FS-SEI*

A typical SEI can be decomposed into two parts: feature embedding and classification. The former is to build the mapping from the sample space to the feature space, that is, $f_{\text{FE}}(\mathbf{W}_{\text{FE}}) : \mathcal{X} \rightarrow \mathcal{F}$, and the latter is to map the features into categories, i.e., $f_{cl}(\mathbf{W}_{cl}) : \mathcal{F} \rightarrow \mathcal{Y}$, where $\mathcal{F}$ is the feature space, $f_{\text{SEI}}(\mathbf{W}) = f_{cl}[f_{\text{FE}}(\mathbf{W}_{\text{FE}}); \mathbf{W}_{cl}]$ and $\mathbf{W} = \{\mathbf{W}_{\text{FE}}, \mathbf{W}_{cl}\}$.

FS-SEI also has the similar implementation. However, it is difficult to construct the feature embedding based on DL model with huge hyper-parameters in a scenario where samples are scarce, while the realization of the classification has almost no difficulties, because there are lots of classifiers with few parameters or even no parameters (For instance, some ML classifier and distance-based classifiers). Thus, the key of FS-SEI is constructing a mapping for a good feature embedding, i.e., $f_{\text{FE}}(\mathbf{W}_{\text{FE}}) : \mathcal{X}_{fs} \rightarrow \mathcal{F}$. In addition, considering that the direct construction is rarely possible and $\mathcal{X}_{fs} = \mathcal{X}_{au}$, the general alternative is to construct $f_{\text{FE}}(\mathbf{W}_{\text{FE}}) : \mathcal{X}_{au} \rightarrow \mathcal{F}$.

In this paper, the feature embedding is realized by CVCNN and metric learning, while the classification is based on ensemble learning. The overview of the proposed deep metric ensemble learning (DMEL)-based FS-SEI method is shown in Fig. 3.

### *B. Feature Embedding via CVCNN and Metric Learning*

*1) The structure of CVCNN for SEI:* In this paper, CVCNN works as the feature embedding $f_{\text{FE}}(\mathbf{W}_{\text{FE}})$ that is used to map ADS-B signal samples to features, because it has much more powerful and effective feature representation and classification capability than real-valued CNN

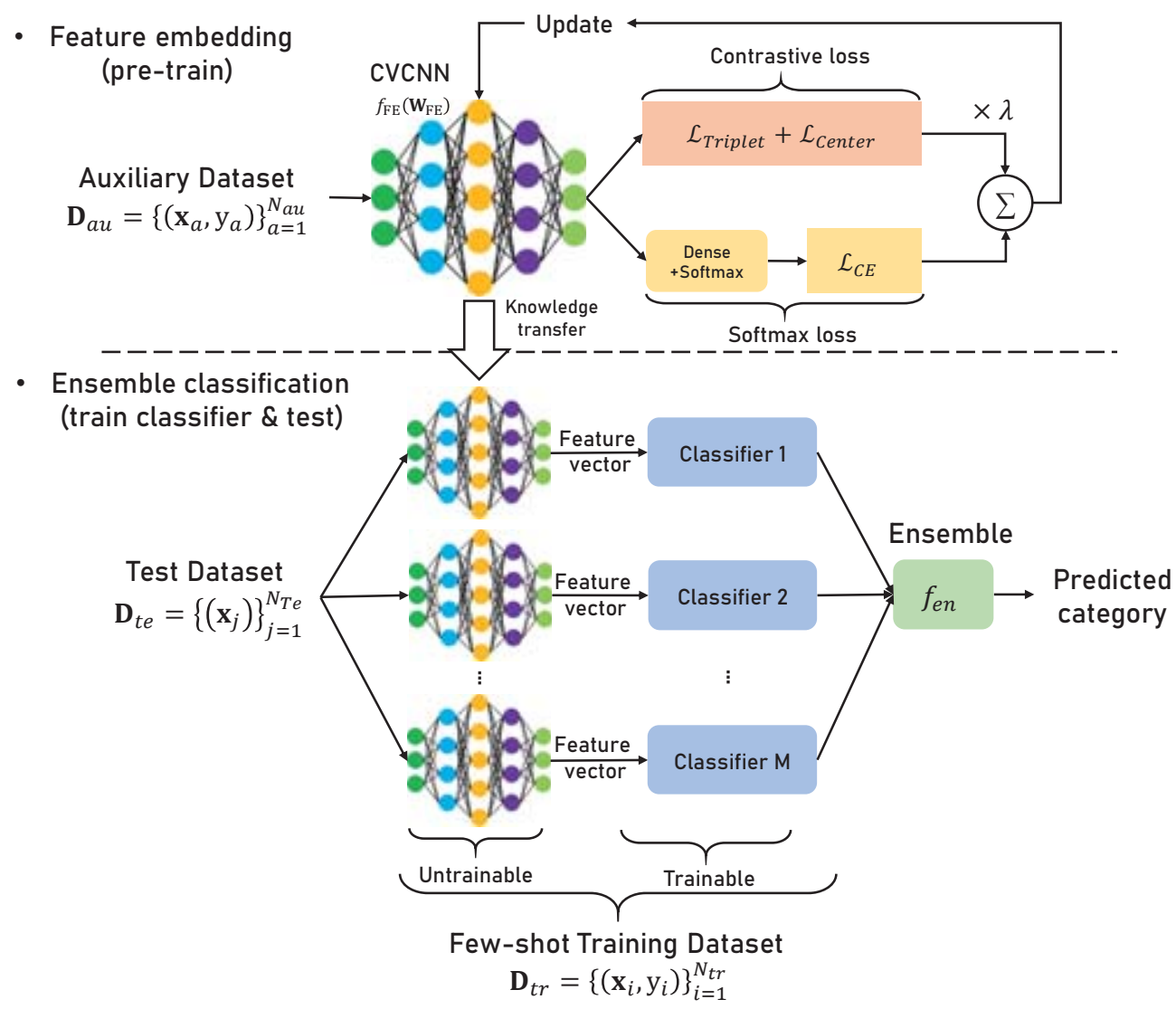

Fig. 3. The overview of FS-SEI based on DMEL.

TABLE II

THE STRUCTURE OF CVNN WITH MULTIPLE COMPLEX-VALUED CONVOLUTIONAL LAYERS AND ONE FULLY-CONNECTED LAYER.

| Structure | The number of layers |
|---|---|
| Input (samples, categories) | - |
| CVConv1D ($N_{ne}$, 3) + CVReLU + CVBN + Maxpooling1D (2) | $\times 9$ |
| Flatten | $\times 1$ |
| Dense (1024)+ ReLU + Dropout (0.5) | $\times 1$ |

Note: "$\times$9" represents that there are nine same layers, while the meaning of "$\times$1" is similar.

in signal identification [24], [49]. The specific structure and parameters of the CVCNN are shown in TABLE II. Here, CVCNN consists of nine convolution layers and one fully-connected layer. There are three essential components of CVCNN: complex-valued convolution layer, complex-valued batch normalization layer, and complex-valued activation function, which will be introduced as follows.

**Complex-valued convolution layer:** The input of the CVCNN is a complex-valued matrix with dimensionality $N_{in} \times C_{in}$, and it can be denoted as $\mathbf{I}_{in}$. The complex-valued convolution

operation is give as

$$\begin{aligned}\mathbf{I}_{out} &= \sum_{n=1}^{N_{in}} \mathbf{W}_n * \mathbf{I}_{in}(\cdot, n) \\ &= \sum_{n=1}^{N_{in}} [\Re(\mathbf{W}_n) + j \cdot \Im(\mathbf{W}_n)] \\ &\qquad * [\Re(\mathbf{I}(\cdot, n)) + j \cdot \Im(\mathbf{I}_{in}(\cdot, n))], \\ &= \sum_{n=1}^{N_{in}} \Re(\mathbf{W}_n) * \Re(\mathbf{I}_{in}(\cdot, n)) + j \cdot \Im(\mathbf{W}_n) * \Re(\mathbf{I}_{in}(\cdot, n)) \\ &\quad + j \cdot \Re(\mathbf{W}_n) * \Im(\mathbf{I}_{in}(\cdot, n)) - \Im(\mathbf{W}_n) * \Im(\mathbf{I}_{in}(\cdot, n)),\end{aligned} \tag{6}$$

where $\mathbf{W}_n$ are the complex-valued convolution kernels in the $n$-th channel; '*' represents convolution operation.

The convolution operation for SEI in this paper is one-dimensional convolution, and the dimensionality of convolution kernel is $N_{ke}$. Assuming that there are $N_{ne}$ neurons and the convolution stride is 1, the time complexity of one complex-valued convolution layer can be written as

$$Time \sim O(4N_{ke} \cdot N_{out} \cdot N_{in} \cdot N_{ne}), \tag{7}$$

where $N_{out}$ is equal to $(N_{in} - N_{ke} + 1)$ (the padding mode is "valid") or $N_{in}$ (the padding mode is "same"), respectively.

**Complex-valued batch normalization:** Batch normalization (BN) is generally used to accelerate the training process and avoid overfitting, and there is the corresponding complex-valued BN (CVBN). Assuming the $i$-th element of the complex-valued input in a batch is $I_{b,i}$. Then, the real part and imaginary part of $I_{b,i}$ are split to form a matrix

$$\mathbf{I}_{b,i} = \begin{bmatrix} \Re(I_{b,i}) \\ \Im(I_{b,i}) \end{bmatrix}_{2\times 1}.$$

Similar with the real-valued BN, CVBN can be expressed as

$$f_{CVBN}(\mathbf{I}_{b,i}) = \boldsymbol{\gamma}(\boldsymbol{\sigma}^2 + \boldsymbol{\epsilon})^{-\frac{1}{2}}[\mathbf{I}_{b,i} - \boldsymbol{\mu}] + \boldsymbol{\beta}, \tag{8}$$

where

$$\boldsymbol{\sigma}^2 = \begin{bmatrix} Var_{\Re} & Cov_{\Re\Im} \\ Cov_{\Im\Re} & Var_{\Im} \end{bmatrix},$$

$$\boldsymbol{\mu} = \begin{bmatrix} \frac{1}{B}\sum_{b=1}^{B}\Re(I_{b,i}) \\ \frac{1}{B}\sum_{b=1}^{B}\Im(I_{b,i}) \end{bmatrix},$$

$$\boldsymbol{\gamma} = \begin{bmatrix} \gamma_{\Re\Re} & \gamma_{\Re\Im} \\ \gamma_{\Im\Re} & \gamma_{\Im\Im} \end{bmatrix},$$

$$\boldsymbol{\beta} = \begin{bmatrix} \beta_{\Re} \\ \beta_{\Im} \end{bmatrix},$$

and $(\cdot)^{-1/2}$ is an inverse operation with taking square root; $Cov_{\Re\Im}$ (or $Cov_{\Im\Re}$) is the covariance between $\{\Re(I_{b,i})\}_{b=1}^{B}$ and $\{\Im(I_{b,i})\}_{b=1}^{B}$; $Var_{\Re}$ and $Var_{\Im}$ are the variance of $\{\Re(I_{b,i})\}_{b=1}^{B}$ and $\{\Im(I_{b,i})\}_{b=1}^{B}$, respectively; $\boldsymbol{\beta}$ and $\boldsymbol{\gamma}$ are the learnable parameters; $B$ is the number of batch size.

**Complex-valued activation function:** Complex-valued rectified linear unit (CVReLU) can be expressed as

$$f_{CVReLU}(\mathbf{I}_{in}) = \max[\Re(\mathbf{I}_{in}), 0] + j \cdot \max[\Im(\mathbf{I}_{in}), 0]. \tag{9}$$

*2) Hybrid metric for discriminative feature embedding:* Most of conventional SEI methods are close-set identification method, i.e., the categories of samples for testing are within training dataset. In this case, training DL model by Softmax loss can generally obtain good performance. Therefore, the feature embedding based on Softmax loss can only obtain features that are prone to separate different seen categories.

However, the close-set SEI is nearly impossible in realistic deployment, since that it is almost impractical to pre-collect ADS-B signals samples from all possible aircrafts for training. It means that the separable feature is not effective enough, and the features for SEI need to be discriminative and generalized enough for characterizing the ADS-B signal samples from unseen categories or categories seen a few times. The difference between separable feature and discriminative feature is shown as follows.

The core characteristics of discriminative features are the intra-category consistency and the inter-category separability. Thus, we proposed an effective discriminative feature embedding method based on hybrid metric, which consist of Softmax loss, and two typical contrastive losses, and this method can effectively narrow the distances between different ADS-B signal samples from the same aircrafts in the feature space, meanwhile enlarge the distances between ADS-B signal samples from different aircrafts.

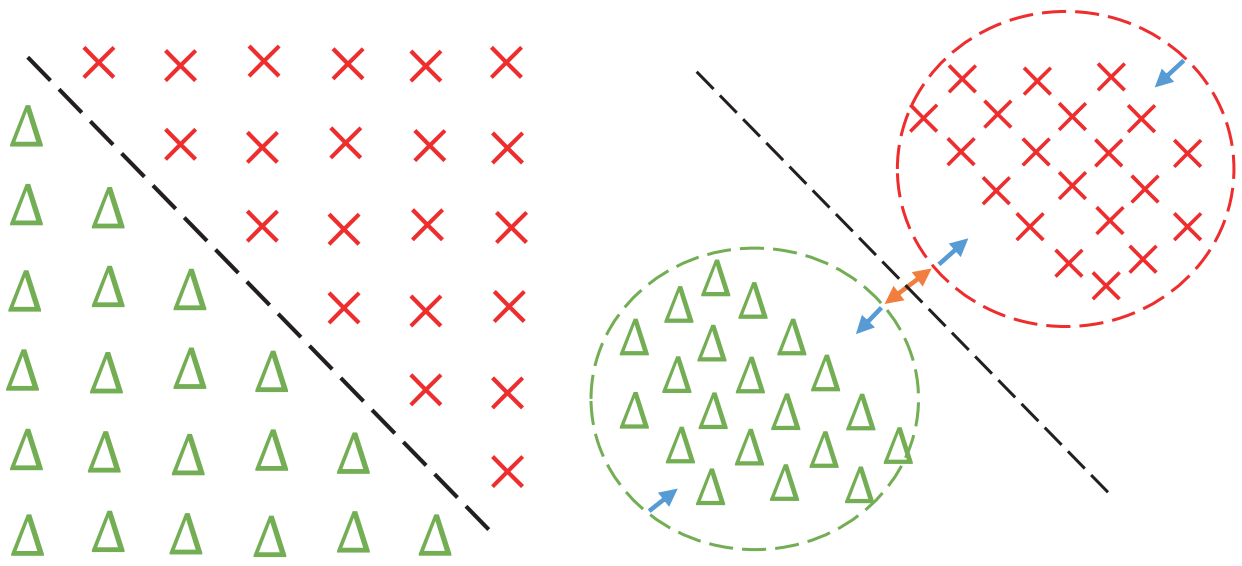

(a) Separable feature (b) Discriminative feature

Fig. 4. The sketch maps of separable feature and discriminative feature. "↔" represents enlarging distance, and "→/←" represents narrowing distance. The relationships between separable feature and discriminative feature is that the discriminative feature is separable, but the separable feature is usually indiscriminative.

**Softmax loss:** The feature, extracted from ADS-B signal sample $\mathbf{x}_a$ in $\mathbf{D}_{au}$, can be expressed as $\mathbf{f}_a = f_{\mathrm{FE}}(\mathbf{W}_{\mathrm{FE}}; \mathbf{x}_a)$. Then, this feature is fed into a Dense layer for mapping from feature space into category space corresponding to $\mathbf{D}_{au}$, and the output can be expressed as $\mathbf{z}_a = f_{\mathrm{De}}[\mathbf{W}_{\mathrm{De}}; \mathbf{f}_a]$, where $\mathbf{W}_{\mathrm{De}}$ is the weight of the Dense layer. Thus, Softmax loss can be written as

$$\mathcal{L}_{Softmax} = -\mathbb{E}\left\{\log\left[\frac{e^{\mathbf{z}_a(\mathrm{y}_a)}}{\sum_n e^{\mathbf{z}_a(n)}}\right]\right\}, \tag{10}$$

where $\mathbf{z}_a(\cdot)$ is the $(\cdot)$-th element, which represents the non-normalized probability that the sample belongs to the $c$-th category. In addition, it is noted that the above function is a simplified form.

**Contrastive loss:** Softmax loss is separable enough, but its generalization and discrimination are insufficient. Therefore, two contrastive losses are introduced as auxiliary losses to supplement these deficiencies. The one of the contrastive losses is triplet loss, which can narrow the intra-category distance and enlarge the inter-category distance in the feature space. Triplet loss, applied in this paper, is originate from triplet network [52], which is shown in Fig. 5.

Triplet network divides samples in the dataset $\mathbf{D}_{au}$ into three parts: anchor samples, positive samples and negative samples, where the positive samples are the samples with the same categories of the anchor samples, and the negative samples are the samples that have different categories with the anchor samples. Three kinds of samples are fed into the CVCNN-based feature embedding in pairs, and triplet network encourages the CVCNN-base feature embedding to shrink the distance between $\mathbf{x}^{an}$ and $\mathbf{x}^{+}$, meanwhile expand the distance between $\mathbf{x}^{an}$ and $\mathbf{x}^{-}$ in the feature space. Here, Euclidean distance is generally applied to measure the distance

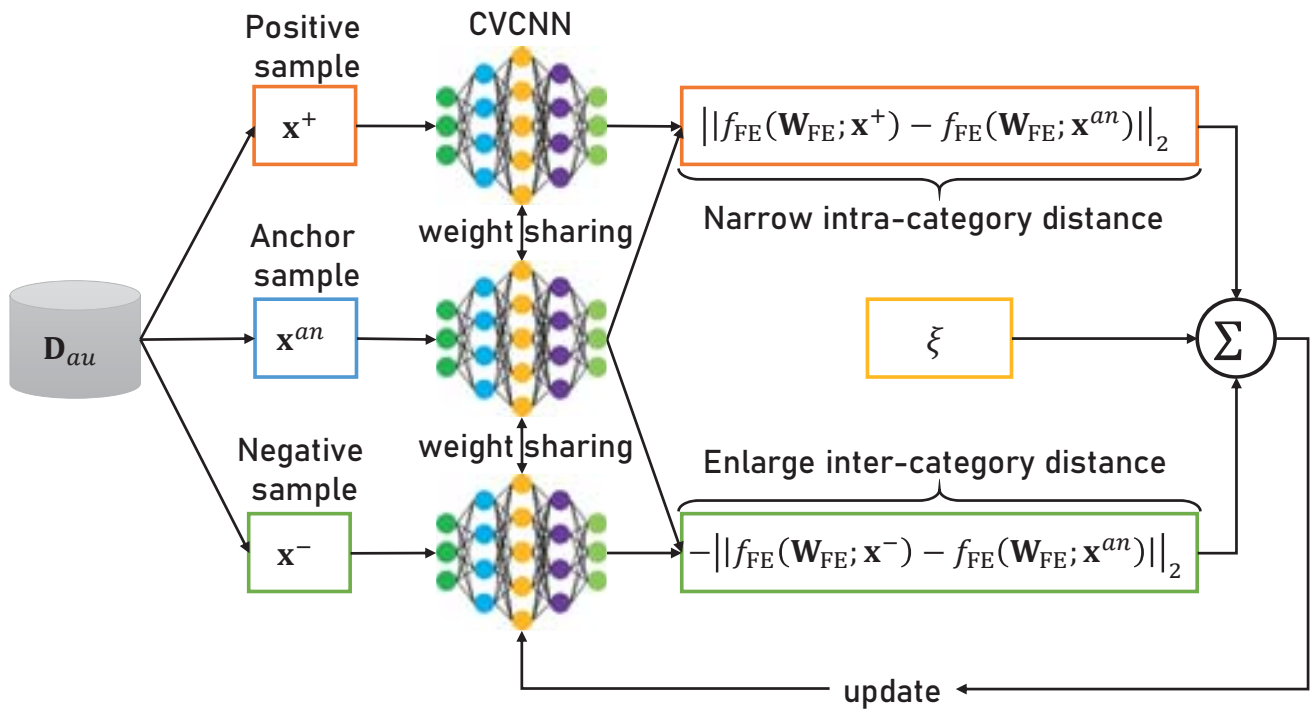

Fig. 5. The principle of triplet loss, which is originate from triplet network.

in the feature space. So, triplet loss can be expressed as

$$\begin{aligned}\mathcal{L}_{Triplet} =&\mathbb{E}\{\max[||f_{\mathrm{FE}}(\mathbf{W}_{\mathrm{FE}};\mathbf{x}^{+}) - f_{\mathrm{FE}}(\mathbf{W}_{\mathrm{FE}};\mathbf{x}^{an})||_2 \\ &- ||f_{\mathrm{FE}}(\mathbf{W}_{\mathrm{FE}};\mathbf{x}^{-}) - f_{\mathrm{FE}}(\mathbf{W}_{\mathrm{FE}};\mathbf{x}^{an})||_2 + \xi, 0]\},\end{aligned} \tag{11}$$

where $||\cdot||_2$ is the $\ell_2$ norm and $\xi$ is the tunable margin.

The other one of the contrastive losses is center loss [53], which works for obtaining more compact intra-category distances in the feature space. Its formula can be written as

$$\mathcal{L}_{Center} = \frac{1}{2}\mathbb{E}\left[||f_{\mathrm{FE}}(\mathbf{W}_{\mathrm{FE}};\mathbf{x}_a) - \mathbf{c}_{\mathrm{y}_a}||_2^2\right], \tag{12}$$

where $\mathbf{c}_{\mathrm{y}_a}$ is the learnable center feature of the $\mathrm{y}_a$-th category. Moreover, the sketch maps of triplet loss and center loss are shown as follows.

Finally, we give the hybrid metric-based loss function for discriminative feature embedding in SEI, which is shown as follows.

$$\begin{aligned}\mathcal{L}_{\mathrm{HM}} &= \mathcal{L}_{Softmax} + \lambda\mathcal{L}_{Contrastive}, \\ &= \mathcal{L}_{Softmax} + \lambda(\mathcal{L}_{Triplet} + \mathcal{L}_{Center}),\end{aligned} \tag{13}$$

where $\lambda$ is a factor to balance Softmax loss and contrastive loss.

*3) Optimization methods:* There are two parts that require to be optimized: the weight of feature embedding $\mathbf{W}_{\mathrm{FE}}$, and the center features $\mathbf{c}_{\mathrm{y}_a}$. The specific optimization methods are shown as follows.

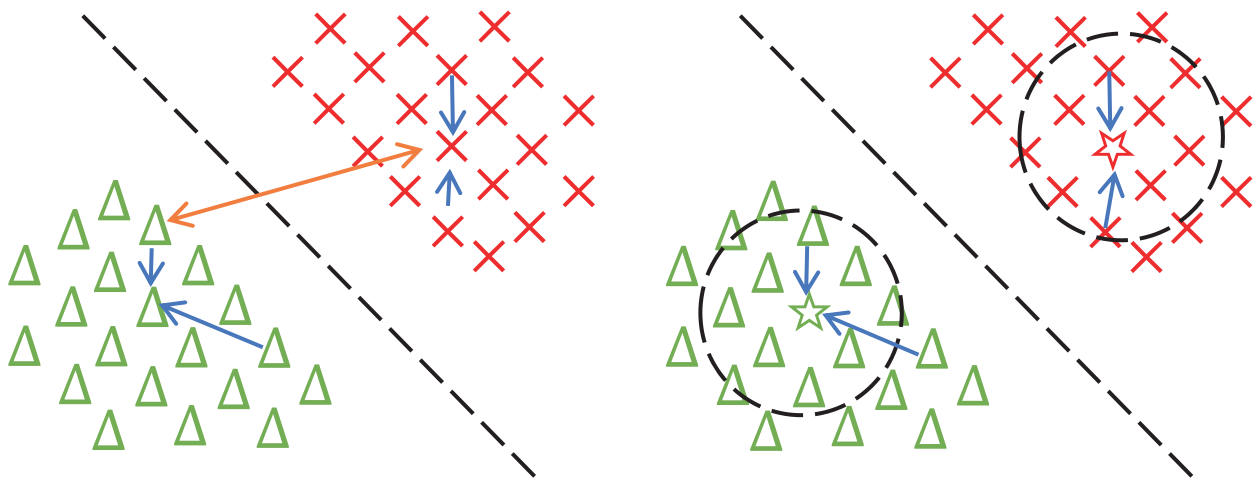

(a) Triplet loss (b) Center loss

Fig. 6. The sketch maps of triplet loss and center loss. "↔" represents enlarging distance, and "→/←" represents narrowing distance.

**The optimization method for feature embedding:** The weight of feature embedding $\mathbf{W}_{\mathrm{FE}}$ is updated by stochastic gradient descent algorithm, which can be written as follows.

$$\mathbf{W}_{\mathrm{FE}}^{t+1} = \mathbf{W}_{\mathrm{FE}}^{t} - \eta \cdot \frac{\partial \mathcal{L}_{\mathrm{HM}}}{\partial \mathbf{W}_{\mathrm{FE}}}, \tag{14}$$

where $\eta$ is the learning rate in SGD.

**The optimization method for center feature:** In the ideal case, if the features are changed, $\mathbf{c}_{\mathrm{y}_a}$ should be updated. However, it is impractical to take entire auxiliary dataset into account, and calculate the average feature of each category. Thus, the center features are updated based on mini-batch method, and it means that the feature centers are calculated by averaging the features of the categories containing in the mini-batch. The update formula of $\mathbf{c}_{\mathrm{y}_a}$ can be written as

$$\mathbf{c}_{\mathrm{y}_a}^{t+1} = \mathbf{c}_{\mathrm{y}_a}^{t} - \alpha \cdot \Delta \mathbf{c}_{\mathrm{y}_a}, \tag{15}$$

$$\Delta \mathbf{c}_{\mathrm{y}_a} = \frac{\sum_{b=1}^{B} \{\delta(\mathrm{y}_b == \mathrm{y}_a) \cdot [f_{\mathrm{FE}}(\mathbf{W}_{\mathrm{FE}}; \mathbf{x}_b) - \mathbf{c}_{\mathrm{y}_a}]\}}{1 + \sum_{b=1}^{B} \delta(\mathrm{y}_b == \mathrm{y}_a)}, \tag{16}$$

where $\alpha$ is the learning rate, and $\delta(\cdot)$ is a conditional selection function. If the condition is established, $\delta(\cdot)$ will be equal to 1, otherwise it will be 0. In other world, if there are no samples corresponding to the $\mathrm{y}_a$-th category in a mini-batch dataset $\{\mathbf{x}_b, \mathrm{y}_b\}_{b=1}^{B}$, the feature center of the $\mathrm{y}_a$-th category remains unchanged.

### *C. Ensemble Classier Based on Probability Average Method*

As mentioned above, FS-SEI can be described as the cascaded feature embedding and a simple classifier, i.e., $f_{\mathrm{SEI}}(\mathbf{W}) = f_{cl}[f_{\mathrm{FE}}(\mathbf{W}_{\mathrm{FE}}); \mathbf{W}_{cl}]$. Thus, after the implement of the hybrid metric-based feature embedding $f_{\mathrm{FE}}(\mathbf{W}_{\mathrm{FE}})$, a simple classifier $f_{cl}(\mathbf{W}_{cl})$ is applied for FS-SEI. Here,

**Algorithm 1** Feature embedding based on CVCNN and metric learning.

**I**nput: Auxiliary dataset $\mathbf{D}_{au}$;

**O**utput: $f_{\mathrm{FE}}(\mathbf{W}_{\mathrm{FE}})$;

Construct CVCNN by TABLE II for feature embedding $f_{\mathrm{FE}}(\mathbf{W}_{\mathrm{FE}})$;

Add two contrastive losses $\mathcal{L}_{Center}$ and $\mathcal{L}_{Triplet}$;

Insert one Dense layer following behind CVCNN $f_{\mathrm{De}}[\mathbf{W}_{\mathrm{De}}; f_{\mathrm{FE}}(\mathbf{W}_{\mathrm{FE}})]$

Add Softmax loss $\mathcal{L}_{Softmax}$;

**Set hyperparameters**:

- Set learning rate $\alpha$ and $\eta$;
- Set maximum epochs $T$;
- Set suitable $\lambda$ and $\xi$;

**Train** on $\mathbf{D}_{au} = \{\mathbf{x}_a, \mathrm{y}_a\}_{a=1}^{N_{au}}$:

**for** $t = 1$ to $T$ **do**:

  **Forward propagation**:

    $\mathbf{f}_a = f_{\mathrm{FE}}(\mathbf{W}_{\mathrm{FE}}^t; \mathbf{x}_a)$;

    $\mathbf{z}_a = f_{\mathrm{De}}(\mathbf{W}_{\mathrm{De}}^t; \mathbf{f}_a)$;

  **Backward propagation**:

    Update $\mathbf{W}_{\mathrm{FE}}^t$ by (14)

    Update $\mathbf{c}_{\mathrm{y}_a}^t$ by (15)

    Update $\mathbf{W}_{\mathrm{De}}^{t+1} = \mathbf{W}_{\mathrm{De}}^t - \eta \cdot \frac{\partial \mathcal{L}_{\mathrm{Softmax}}}{\partial \mathbf{W}_{\mathrm{De}}}$

**end for**

**return** $f_{\mathrm{FE}}(\mathbf{W}_{\mathrm{FE}})$

taking logistic regression (LR) as an example, the training and testing processing of classifier is given as follows.

The training process is based on the few-shot ADS-B signal training dataset $\mathbf{D}_{tr}$, which can be described as follows.

$$\mathbf{W}_{cl}^{*} = \arg\min_{\mathbf{W}_{cl}} \left\{ -\frac{1}{N_{tr}} \sum_{i=1}^{N_{tr}} \log \mathbf{P}_i(\mathrm{y}_i) \right\}, \tag{17}$$

where $\mathbf{P}_i = f_{\mathrm{SEI}}(\mathbf{x}_i; \mathbf{W})$ is the predicted probability distribution about $\mathbf{x}_i$, where $\mathbf{x}_i \in \mathbf{D}_{tr}$, and $\mathbf{P}_i(\mathrm{y}_i)$ is the $\mathrm{y}_i$-th element, which represents the predicted probability of the the $\mathrm{y}_i$-th categories.

Next, the test process on $\mathbf{D}_{te}$ can be written as

$$\hat{\mathrm{y}}_j = \underset{1\leq \mathrm{y}_j \leq \mathrm{C}}{\arg\max}\, \mathbf{P}_j, \tag{18}$$

where $\hat{\mathrm{y}}_j$ is the predicted category, and $\mathbf{P}_j = f_{cl}[f_{\mathrm{FE}}(\mathbf{x}_j; \mathbf{W}_{\mathrm{FE}}); \mathbf{W}_{cl}]$, $\mathbf{x}_j \in \mathbf{D}_{te}$. This process means that the category with the highest probability is as the predicted category.

To further improve the identification performance and robustness, ensemble learning is introduced, which consists of two steps: construction of multiple base FS-SEI models with diversity and joint decision. In the former step, base SEI models are built by multiple independent training for diversity. These model weights can be written as $\{\mathbf{W}^m\}_{m=1}^M$, where $M$ is the number of base FS-SEI models. The latter step is based on probability average method, which can be expressed as

$$\hat{\mathrm{y}}_j^{en} = f_{en}(\mathbf{P}_j^1, \mathbf{P}_j^2, \cdots, \mathbf{P}_j^M) = \underset{1\leq \mathrm{y}_j \leq \mathrm{C}}{\arg\max} \frac{1}{M}\sum_{m=1}^{M} \mathbf{P}_j^m, \tag{19}$$

where $\hat{\mathrm{y}}_j^{en}$ is the predicted category by ensemble learning, and $\mathbf{P}_j^m = f_{\mathrm{SEI}}(\mathbf{x}_j; \mathbf{W}^m)$, $\mathbf{x}_j \in \mathbf{D}_{te}$.

**Algorithm 2** Identification method via ensemble learning.

**I**nput: Few shot training dataset $\mathbf{D}_{tr}$ and test dataset $\mathbf{D}_{te}$;

**O**utput: Predicted results $\{\hat{\mathrm{y}}_j^{en}\}_{j=1}^{N_{te}}$;

1: Set the number of base classifiers $M$;
2: Obtain $\mathbf{W}_{\mathrm{FE}}^m$, $m = 1, 2, \cdots, M$ by multiple independent model initialization and training;
**Train** on $\mathbf{D}_{tr} = \{\mathbf{x}_i, \mathrm{y}_i\}_{i=1}^{N_{tr}}$:
3: **for** $m = 1$ to $M$ **do**:
4: Add LR classifier $f_{cl}$;
5: Obtain $\mathbf{f}_i^m = f_{\mathrm{FE}}(\mathbf{x}_i; \mathbf{W}_{\mathrm{FE}}^m)$, $i \in \{1, 2, \cdots, N_{tr}\}$;
6: Obtain $\mathbf{W}_{cl}^m$ based on $\mathbf{f}_i^m$ and $\mathrm{y}_i$ by minimizing (17);
7: **end for**
**Test** on $\mathbf{D}_{te} = \{\mathbf{x}_j\}_{i=1}^{N_{te}}$:
8: Cascade $f_{\mathrm{FE}}$ and $f_{cl}$, i.e., $f_{\mathrm{SEI}}(\mathbf{W}^m) = f_{cl}[f_{\mathrm{FE}}(\mathbf{W}_{\mathrm{FE}}^m); \mathbf{W}_{cl}^m]$;
9: Calculate the predicted probability distribution $\mathbf{P}_j^m = f_{\mathrm{SEI}}(\mathbf{x}_j; \mathbf{W}^m)$;
10: Calculate the the predicted category $\hat{\mathrm{y}}_j^{en}$ by (19);
11: **return** $\hat{\mathrm{y}}_j^{en}$

The training of classifier is completed online. It is noted that the online training and deployment can be realized quickly, because the classifiers for FS-SEI can be few-parameter and even parameterless, and the number of training samples is limited.

### *D. Comparison Methods*

In this paper, the proposed FS-SEI method based on Softmax loss, triplet loss and center loss is named as "STC CVCNN", and other SEI methods for comparison are introduced as follows.

*1) Instantaneous feature:* Instantaneous features are extracted from the amplitude, phase and frequency components of the received signal, and the features include the mean, variance, skewness, and kurtosis [33], [39].

*2) Softmax CVCNN and direct CVCNN:* The CVCNN with only Softmax loss, based on pe-training on $\mathbf{D}_{au}$ and finetuning on $\mathbf{D}_{tr}$, is named as "Softmax CVCNN", and it, based on direct training on $\mathbf{D}_{tr}$, is named as "Direct CVCNN".

*3) Two contrastive loss-based methods:* Here, we adopt two contrastive loss-based signal recognition methods for comparison. The one of the contrastive losses is originate from SiameseNet [44]–[47], which has been introduced in Part II. The other is triplet loss, which has been specifically introduced above. It is noted that the network structures in SiameseNet and TCNN are replaced by the structure in TABLE II for fair comparison. So, they are named as "Siamese CVCNN" and "Triplet CVCNN", and their training schemes are similar with Softmax CVCNN.

*4) SR2CNN:* SR2CNN is an advanced zero-shot signal recognition based on combined metric and generative methods, and it has dual branch structure, respectively responsible for classification and reconstruction. Mean square error is applied as loss function for the reconstruction part, while Softmax loss with center loss is used for classification. It is also a hybrid metric method for signal recognition. Similarly, for fair comparison, the CNN structure in SR2CNN is also replaced with the structure in TABLE II.

*5) Ablation studies:* In addition to the above compared methods, there are two ablation studies, i.e., "ST CVCNN" and "SC CVCNN". The former is the CVCNN based on Softmax loss and triplet loss, while the latter is the CVCNN based on Softmax loss and center loss.

# V. Experimental Results

## A. Simulation Environment, Parameters and Performance Metrics

These DL models are based on Tensorflow, and these machine learning classifiers are based on Scikit-learn. The other simulation parameters are listed in TABLE III. In addition, considering that the samples used in this paper are noisy samples in the real scenarios, and different few-shot training samples have a certain impact on the performance, we randomly selected 1000 groups of different few-shot training samples as $\mathbf{D}_{tr}$ for Monte Carlo simulations, and each simulation is tested on the same test dataset, i.e., $\mathbf{D}_{te}$. The average accuracy of these tests is applied as an indicator of the identification performance of these algorithms.

However, the average accuracy can not directly show the discriminative degrees of different feature embedding methods. So, we also apply the silhouette coefficient (SC) to directly measure the discriminative degree [56]. The SC can measure both cohesion and separability, i.e., the former can indicate the intra-category distance, while the latter can show the inter-category distance. The formula of the SC is given as follows.

$$SC = \frac{1}{N_{te}} \sum_{j=1}^{N_{te}} \frac{D_j^{inter} - D_j^{intra}}{\max(D_j^{inter}, D_j^{intra})}, SC \in [-1, 1], \tag{20}$$

where $D_j^{intra}$ is the average distance between the feature vector of the $j$-th sample and the feature vectors of other samples, which have the same category with the $j$-th sample, i.e.,

$$D_j^{intra} = \frac{\sum_k \delta(\mathrm{y}_j == \mathrm{y}_k) \cdot ||f_{\mathrm{FE}}(\mathbf{x}_j) - f_{\mathrm{FE}}(\mathbf{x}_k)||_2}{\sum_k \delta(\mathrm{y}_j == \mathrm{y}_k)}, \tag{21}$$

where $j \neq k, (\mathbf{x}_j, \mathrm{y}_j) \in \mathbf{D}_{te}$, and $(\mathbf{x}_k, \mathrm{y}_k) \in \mathbf{D}_{te}$. In addition, $D_j^{inter}$ is the minimum distance between the feature vector of the $j$-th sample and the feature vectors of other samples, which have the different category with the $j$-th sample, i.e.,

$$D_j^{inter} = \min[\delta(\mathrm{y}_j \neq \mathrm{y}_k) \cdot ||f_{\mathrm{FE}}(\mathbf{x}_j) - f_{\mathrm{FE}}(\mathbf{x}_k)||_2]. \tag{22}$$

It is obvious that the closer the SC approaches to one, the better the cohesion and separability. In addition, the average accuracy curves are given in Fig. 7∼9, while the SCs are shown in TABLE IV.

TABLE III

THE MAIN PARAMETERS IN THIS PAPER.

| Simulation parameters | Values |
|---|---|
| The dimensionality of samples | $6000 \times 2$ |
| The number of aircrafts in $\mathbf{D}_{au}$ | 90 |
| The number of samples per category in $\mathbf{D}_{au}$ | $200 \sim 500$ |
| The ratio between training and validation samples | 7:3 |
| The number of aircrafts in $\mathbf{D}_{fs}$ | {10, 20, 30} |
| The number of samples per category in $\mathbf{D}_{tr}$ | {1, 5, 10, 15, 20} |
| The number of samples per category in $\mathbf{D}_{te}$ | 200 |
| $N_{ne}$ | 64 |
| Batch size | 32 |
| Maximum epoches | 200 |
| Learning rate $\eta$ and $\alpha$ | 0.001 |
| $\xi$ and $\lambda$ [54], [55] | 5, 0.01 |
| $M$ | {1, 3, 5, 7} |

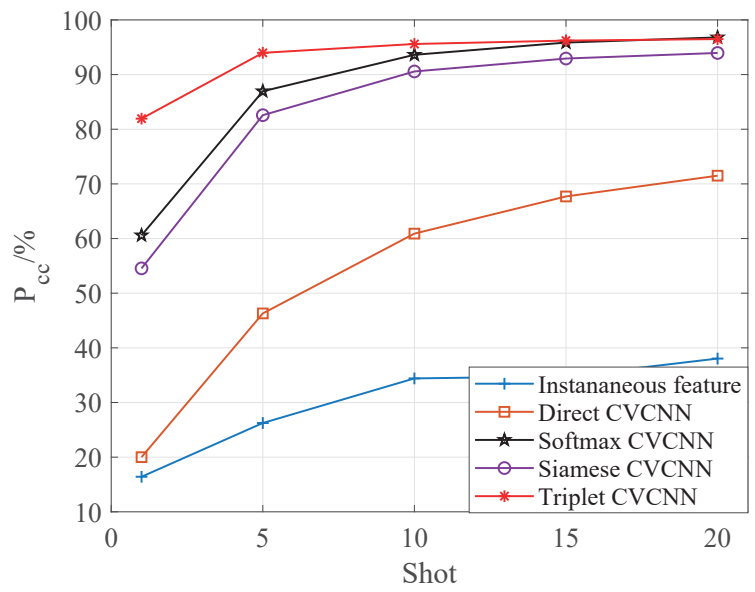

(a) 10 ways

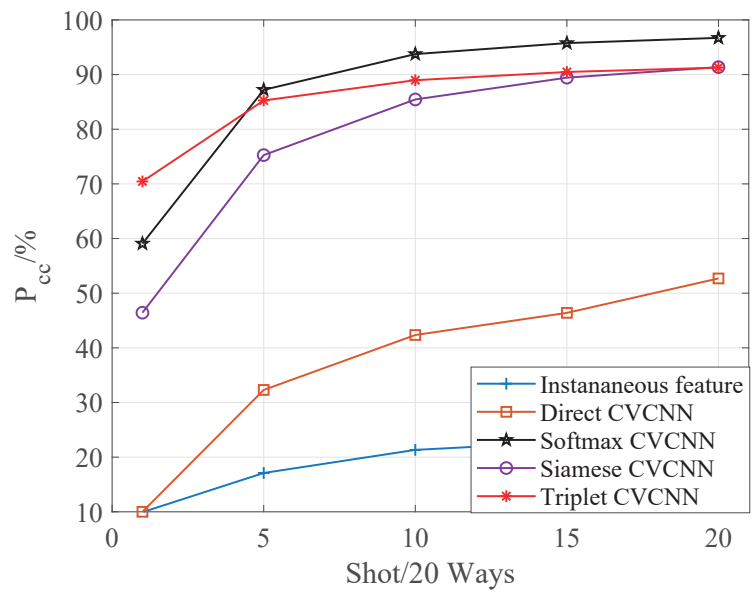

(b) 20 ways

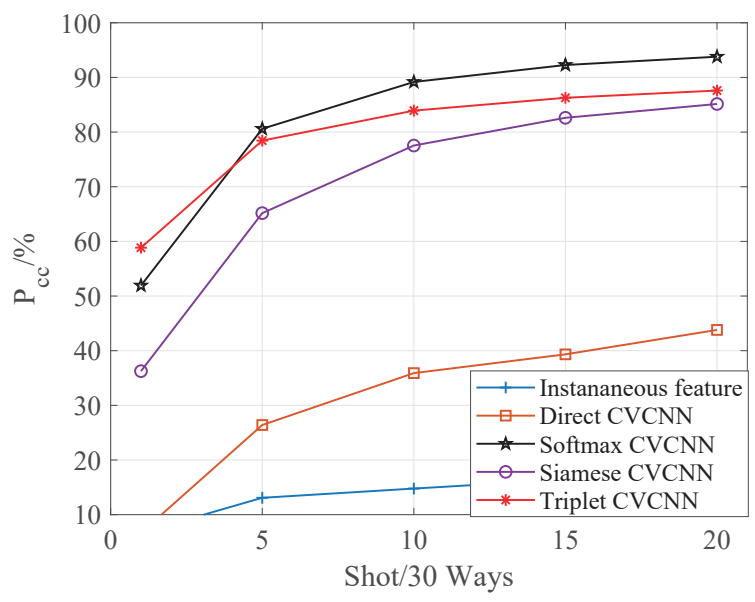

(c) 30 ways

Fig. 7. Softmax loss vs. contrastive loss.

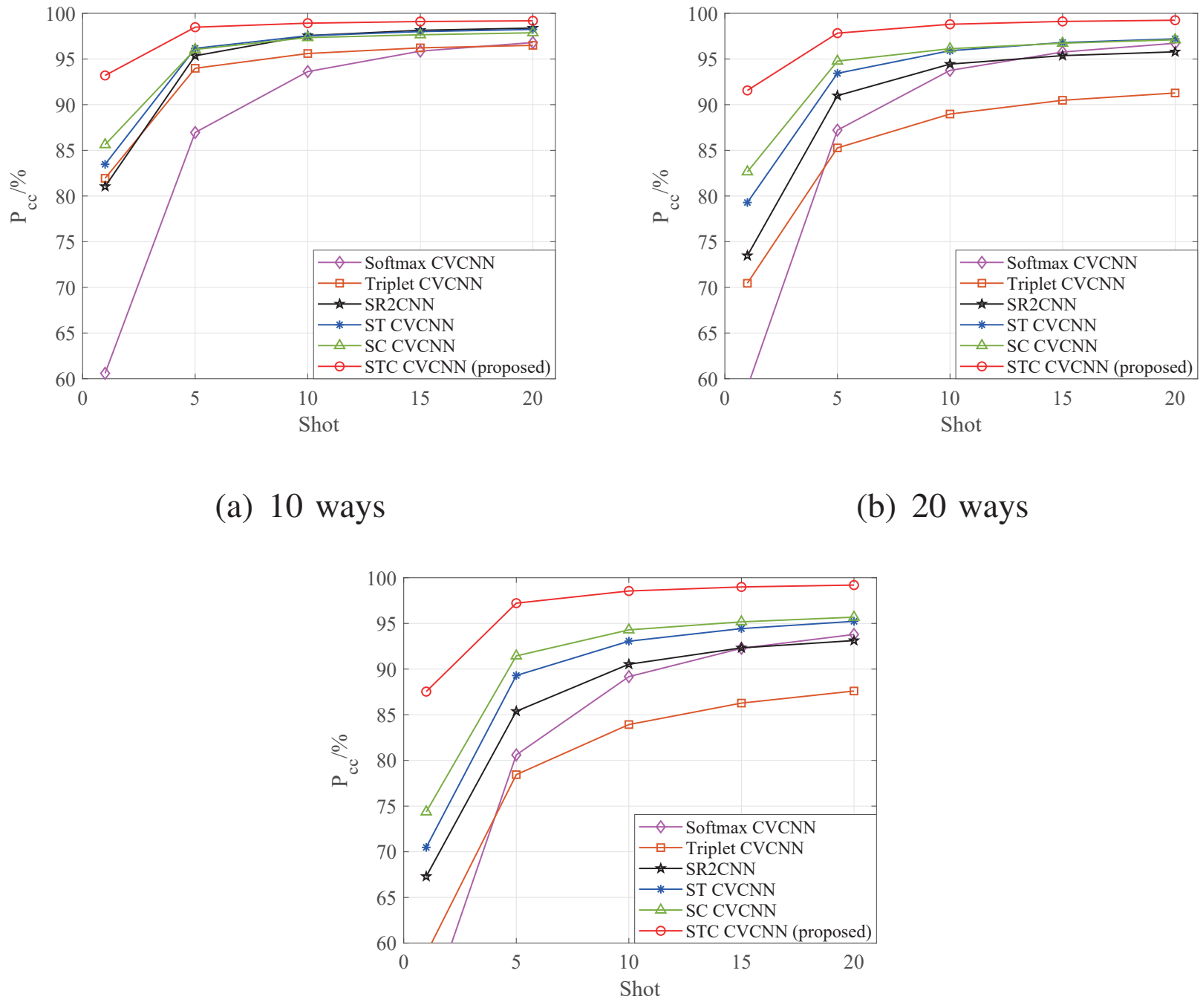

(a) 10 ways (b) 20 ways

(c) 30 ways

Fig. 8. The comparison between different hybrid losses

TABLE IV

THE SCs OF DIFFERENT FEATURE EMBEDDING METHODS.

| ways | Softmax CVCNN | Siamese CVCNN | Triplet CVCNN | SR2CNN | ST CVCNN | SC CVCNN | STC CVCNN (proposed) |
|---|---|---|---|---|---|---|---|
| 10 | 0.0793 | 0.1300 | 0.2537 | 0.2258 | 0.1683 | 0.2846 | **0.4629** |
| 20 | 0.0713 | 0.1300 | 0.2385 | 0.2246 | 0.1663 | 0.2771 | **0.4558** |
| 30 | 0.0588 | 0.0800 | 0.1860 | 0.2092 | 0.1318 | 0.2306 | **0.3722** |

## *B. Softmax Loss vs. Contrastive Loss*

Here, we mainly compare the performance of the CVCNN based on Softmax loss with that based on two contrastive losses. The detailed results are shown in Fig. 7 and TABLE IV.

Firstly, when comparing with two contrastive loss-based methods, we can find that the performance of Triplet CVCNN is obviously better than that of Siamese CVCNN in the average accuracy and the SCs. It can be demonstrated that the former can more effectively narrow the

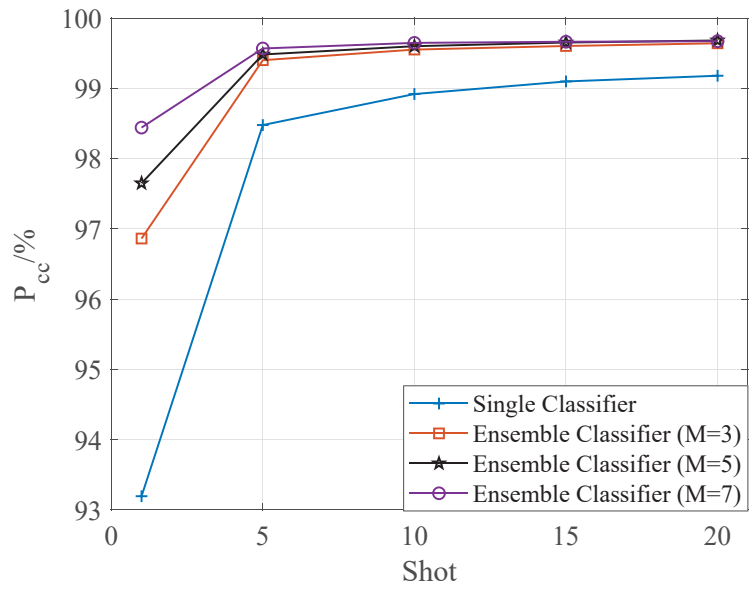

(a) 10 ways

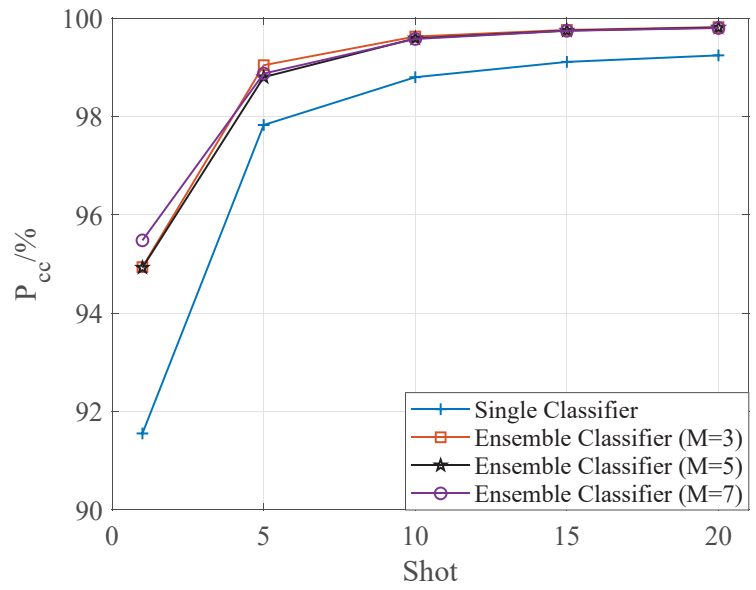

(b) 20 ways

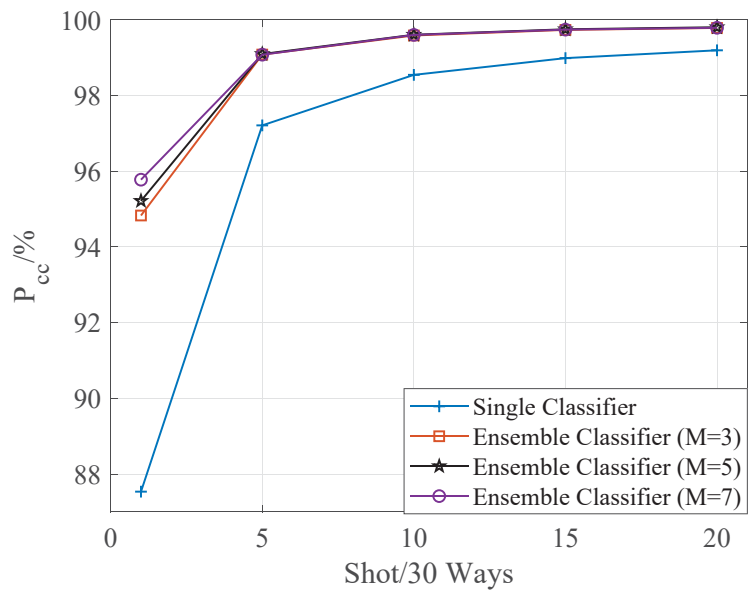

(c) 30 ways

Fig. 9. Single classier vs. ensemble classifier based on STC CVCNN.

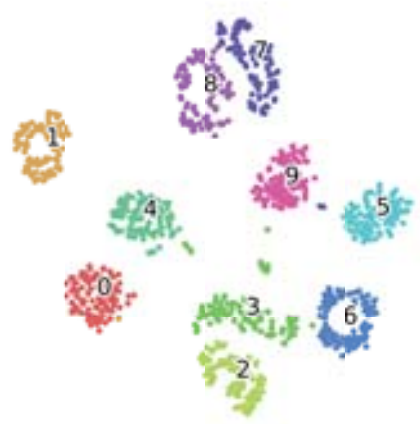

(a) Softmax CVCNN

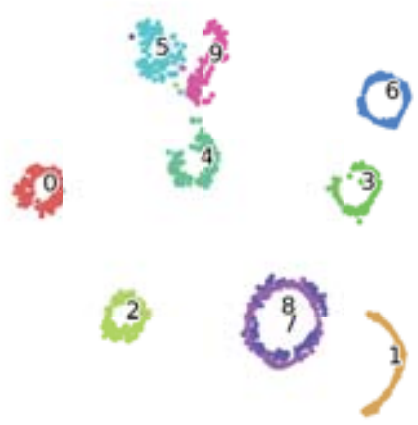

(b) Siamese CVCNN

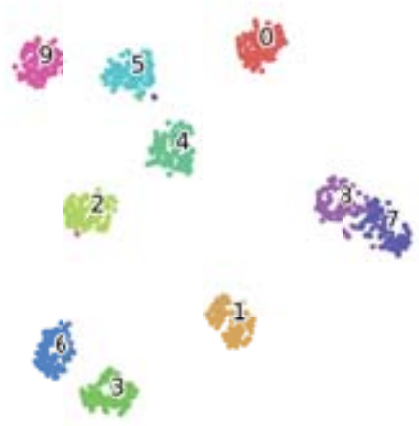

(c) Triplet CVCNN

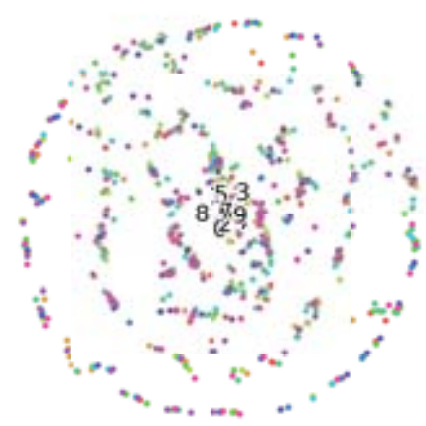

(d) Center CVCNN

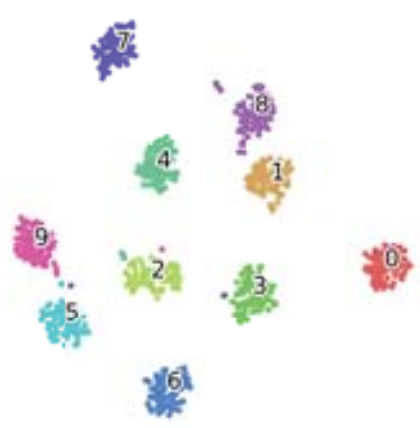

(e) SR2CNN

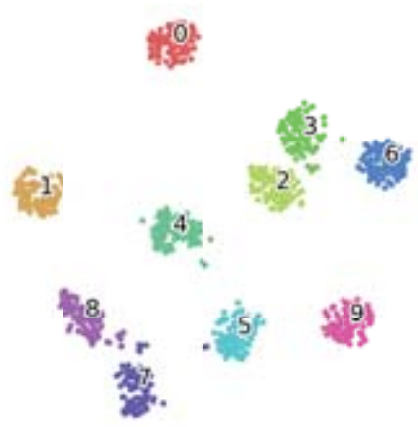

(f) ST CVCNN

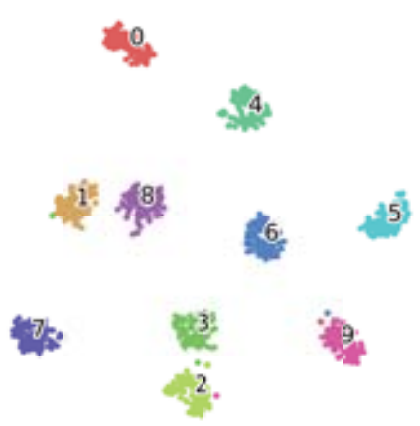

(g) SC CVCNN

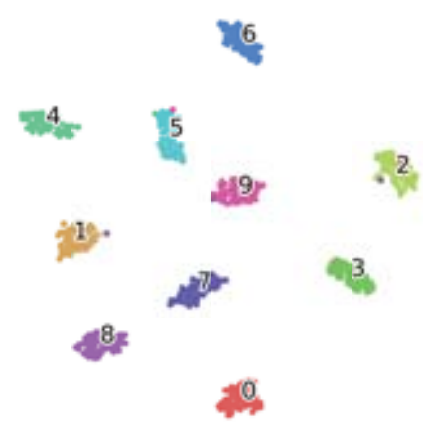

(h) STC CVCNN

Fig. 10. Feature visualization under “10 ways” condition.

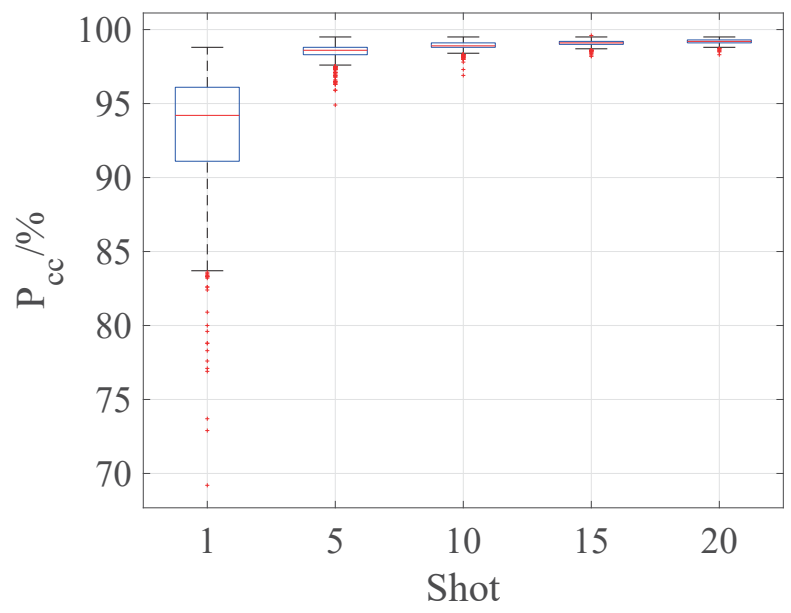

(a) 10 ways

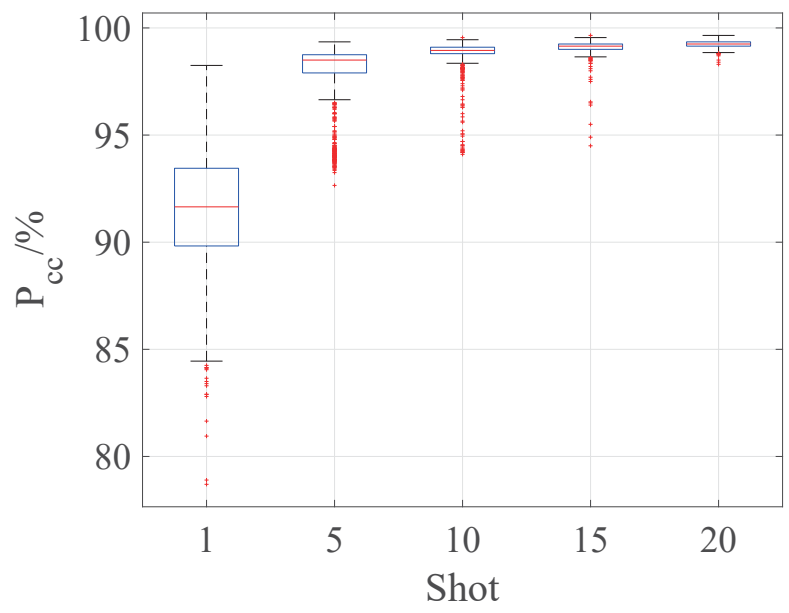

(b) 20 ways

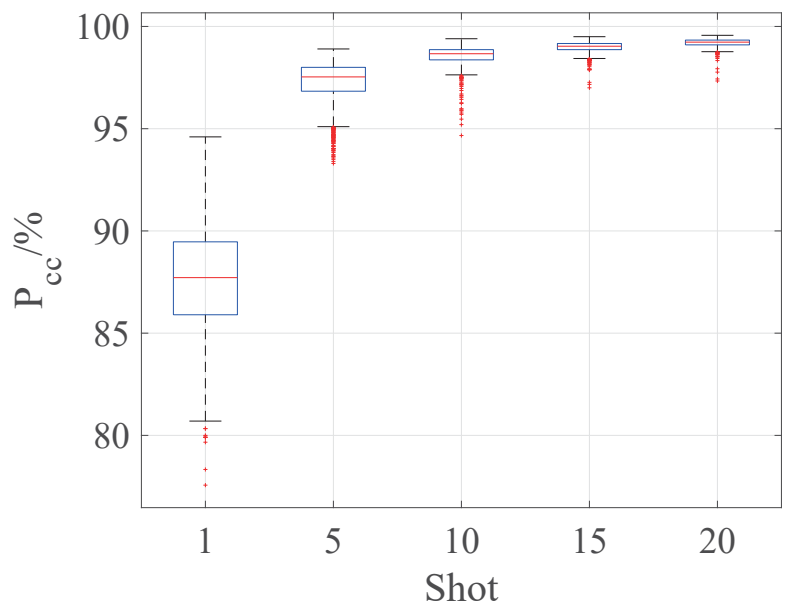

(c) 30 ways

Fig. 11. The box plot of STC CVCNN with single LR classifier.

intra-category distance and enlarge the inter-category distance than the latter, and it is also illustrated by the subsequent feature visualization.

Then, in different cases, Triplet CVCNN and Softmax CVCNN have their own advantages in the average accuracy. Triplet CVCNN generally outperforms Softmax CVCNN in the case of too few training samples (such as 1 shot) or not many aircrafts to be identified (such as 10 ways), but with the increasing of training samples or aircrafts to be identified, the latter has better identification performance than the former. This illustrates that neither of the CVCNN based on single Softmax loss or contrastive loss realizes the sufficient discrimination and generalization, but they have their own advantages. Softmax loss can provide the excellent inter-class separability, while contrastive loss can improve its cohesion. Thus, it may be feasible to combine Softmax loss and contrastive loss for the better discrimination and generalization capabilities, and this is the cornerstone of our subsequent simulation.

Except the above methods with the assistant of auxiliary dataset, the performance Direct

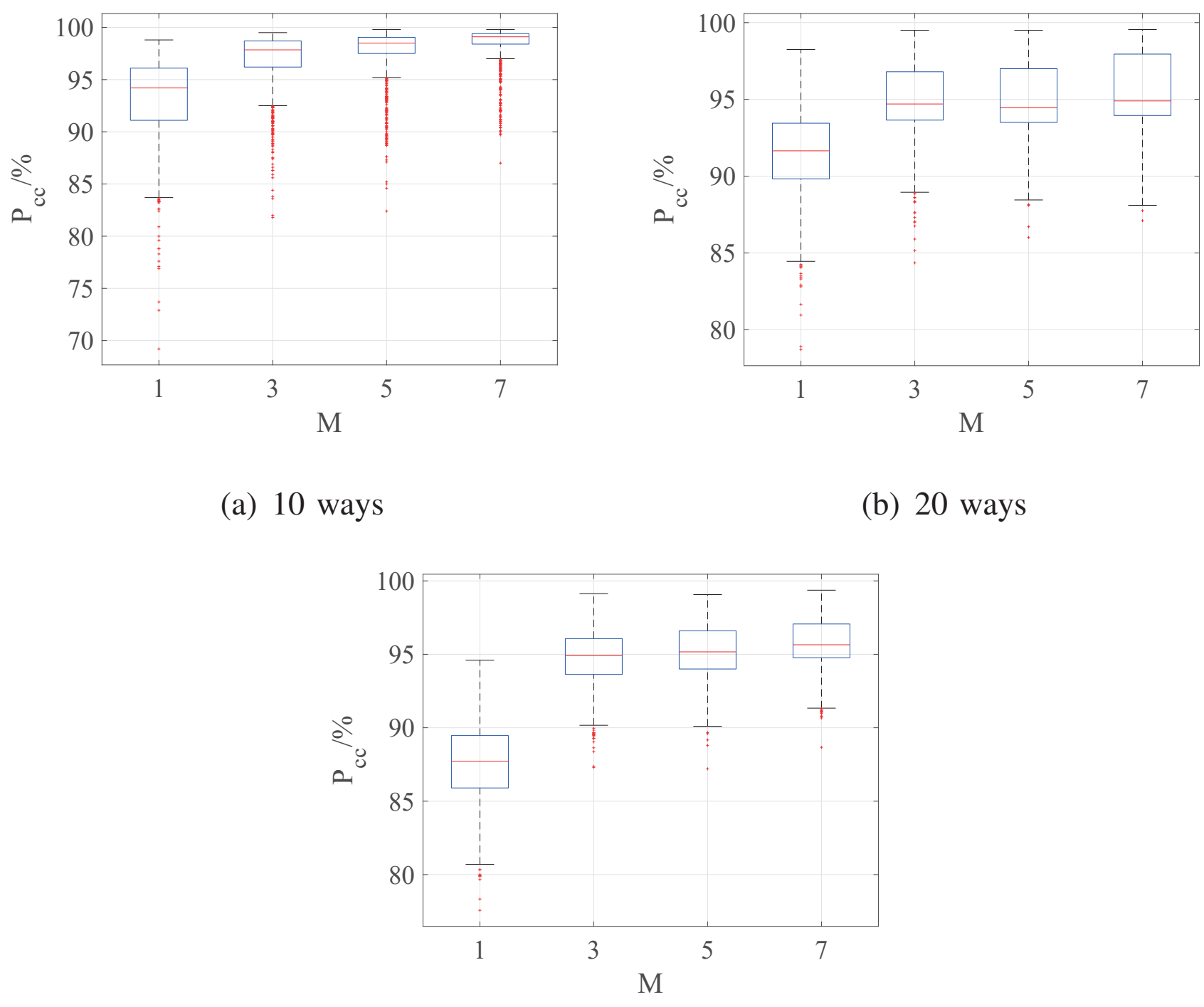

(a) 10 ways

(b) 20 ways

(c) 30 ways

Fig. 12. The box plot of STC CVCNN with ensemble classifier under "1 shot" condition.

CVCNN is also given. Obviously, it has poor identification performances, and its performances are just slightly better than traditional feature-based method. These simulation results reveal the necessity of auxiliary dataset in the FS-SEI problem.

Finally, it is noted that the performance of the center loss-based CVCNN is not given, because feature embedding method, only based on center loss, has little separability, and the identification performance of the center loss-based method is close to that of random guess. It can be demonstrated by feature visualization in Fig. 10(d).

### *C. The Comparison between Different Hybrid Losses*

The performance between different combinations of Softmax loss and contrastive losses is shown in Fig. 8 and TABLE IV.

*1) STC CVCNN vs. SR2CNN vs. single loss-based methods:* Compared with single Softmax loss or contrastive loss, the combination of both has better performance in few-shot scenarios. In most few-shot scenarios, STC CVCNN and SR2CNN outperform single Softmax loss or contrastive loss-based method in the average accuracy. More importantly, our proposed STC CVCNN far exceeds SR2CNN and other methods in the average accuracy and the SCs. Moreover, with the increasing of the number of aircrafts to be identified, the performance gap between STC CVCNN and other methods also increases. It demonstrated that STC CVCNN has stabler performance than other methods under the condition of different ways.

*2) Ablation studies of STC CVCNN:* It is obvious that STC CVCNN outperforms ST CVCNN and SC CVCNN in the average accuracy and the SCs. Specifically, compared with ST CVCNN and SC CVCNN, STC CVCNN has 0.9%∼9.7% and 1.3%∼7.6% identification performance improvement in "10 ways" case, respectively. Furthermore, with the increasing of the number of aircrafts to be identified, their identification performance gaps become larger and larger. Similarly, the SCs of STC CVCNN is higher than that of ST CVCNN and SC CVCNN, which demonstrates that STC CVCNN has better separability and cohesion than other methods.

### *D. Single Classifier vs. Ensemble Classifier*

The above simulation results are based on single machine learning classifier, i.e., LR, and the following ensemble classifier is also applied LR as the base classifier. The detailed results between single classifier and ensemble classifier are shown in Fig. 9. Obviously, regardless of the number of base classifiers, ensemble classifiers perform better than single classifier. In detail, under one shot condition, the performance improvement can exceed 3%∼7%, while that is only 0.5%∼1%. With the increasing of base classifiers, the performance of ensemble classifier can be slightly improved only in case of one shot, and the maximum improvements are 1.58%, 0.54% and 0.95%, but the performances in other cases are barely modified. The above analysis illustrates the effectiveness of ensemble learning. Meanwhile, it also shows that the increasing of base classifiers will not bring much improvement, but only increase the cost of training and deployment, except in the one-shot scenario.

### *E. Feature Visualization*

The dimensionality of the extracted features is reduced to two dimensions by t-distributed stochastic neighbor embedding (t-SNE) for visualization, which is shown in Fig. 10. Here, we only give the scenario of "10 ways", because if there are too many categories, the figure will be too complex and difficult to analyze.

These single loss based methods are shown in Fig. 10(a)∼(c). It is obvious that the feature boundary of each category can be easily determined in the feature space from Softmax CVCNN, but these features of each category are loosely distributed, and do not have compact intra-distance. Instead, Siamese CVCNN and Triplet CVCNN can significantly reduce the intra-distance, especially the latter, but their extracted features have poor separability, for instance, “category 7” and “category 8” almost overlap in Siamese CVCNN, and they have no obvious separation boundary in Triplet CVCNN.

Thus, we combine Softmax loss and contrastive loss to extract the discriminative features. The visualization of these multi-loss based methods is shown in Fig. 10(e)∼(h). Compared with Triplet CVCNN, ST CVCNN only adds extra Softmax loss, and it not only maintains compact intra-category distance, but also slightly enlarge inter-category distance to realize the separability of features. SR2CNN and SC CVCNN also have the similar visualization results.

More than that, STC CVCNN adds extra center loss on the basis of ST CVCNN for further narrowing intra-category distance. It can be observed in Fig. 10(h) that the intra-category distance of features extracted by STC CVCNN is more compact than that extracted by other methods, and STC CVCNN realizes the excellent discriminability of features, which is the key to its performance far surpassing other methods.

## *F. Existing Problem and Analysis*

Here, the detailed analysis of 1000 Monte Carlo simulations is adopted by box plot, which is shown in Fig. 11∼12. The red “+” represents the outlier, and the value of the red horizontal line is median; two black horizontal lines represent the maximum and minimum values in the case of removing outliers, respectively; the upper horizontal line of the blue box is the upper quartile, while the lower one represents the lower quartile.

Fig. 11 revealed the existing problem that different few-shot training samples have different identification performances. A group of excellent few-shot training samples can realize more than 90% accuracy, while the identification performance based on a set of poor quality samples just has 60%∼70%. This phenomenon is the most serious in the one-shot scenario, and the more training samples, the slighter the performance fluctuation. For example, the identification performance ranges from 69.2% to 98.8% under the “1 shot, 10 ways” condition, while the fluctuation range of accuracy is 98.3%∼99.5% under the “20 shot, 10 ways” condition. The possible reason for the above phenomena is that the training dataset is noisy dataset collected from the real open world, and the sample quality is inconsistent. The sample quality may be related to wireless channel, noise, interference, even wrong category caused by wrong demodulation.

The above analysis is based on single LR classifier, and we also give the detail results about ensemble classifier in Fig. 12. Except improving the performance, another original intention of using ensemble classifier is to reduce the effect of sample quality on the stability of identification performance. Excluding a few outliers, ensemble classifier can obviously reduce the performance fluctuation in the scenario of 10 ways, and more based classifiers, more stable identification performance. However, ensemble classifier has little effect on improving the stability in other scenarios.

## VI. Conclusion

In this paper, we proposed an effective FS-SEI method for aircraft identification based on metric learning and ensemble learning. Specifically, the proposed FS-SEI method consists of feature embedding and ensemble classifier. The former is to map from ADS-B signal samples into features by CVCNN and hybrid metric loss combined with Softmax loss and two contrastive losses, while the latter is to construct the mapping from features into categories by a simple ensemble classifier. Simulation results demonstrated the effectiveness of our proposed FS-SEI via metric learning and CVCNN-based feature embedding and ensemble classifier. Feature visualization also showed the compact intra-category distance and separable inter-category distance in the features extracted by our proposed method. Finally, we also revealed the impact of noisy samples on the stability of the algorithm, and we expect to use some schemes, such as attention mechanism [57], to reduce the impact of sample quality on identification performance in the future works.

## References

[1] H. Guo, J. Li, J. Liu, N. Tian and N. Kato, “A survey on space-air-ground-sea integrated network security in 6G,” *IEEE Communications Surveys & Tutorials*, early access, doi: 10.1109/COMST.2021.3131332.

[2] X. Sun, D. W. K. Ng, Z. Ding, Y. Xu and Z. Zhong,“Physical layer security in UAV systems: Challenges and opportunities” *IEEE Wireless Communications*, vol. 26, no. 5, pp. 40–47, Oct. 2019.

[3] G. Gui, M. Liu, F. Tang, N. Kato and F. Adachi, “6G: Opening new horizons for integration of comfort, security, and intelligence” *IEEE Wireless Communications*, vol. 27, no. 5, pp. 126–132, Oct. 2020.

[4] L. Bai, L. Zhu, J. Liu, J. Choi and W. Zhang,"Physical layer authentication in wireless communication networks: A survey" *Journal of Communications and Information Networks*, vol. 5, no. 3, pp. 237–264, Sept. 2020.

[5] N. Wang, W. Li, P. Wang, A. Alipour-Fanid, L. Jiao and K. Zeng, "Physical layer authentication for 5G communications: opportunities and road ahead" *IEEE Network*, vol. 34, no. 6, pp. 198–204, Nov. 2020.

[6] C. Mitchell and C. He, "Security analysis and improvements for IEEE 802.11 i," in *The 12th Annual Network and Distributed System Security Symposium*, 2005, 90–110.

[7] X. Liu, L. Zheng, *et al*., "A broad learning-based comprehensive defence against SSDP reflection attacks in IoTs," *Digital Communications and Networks*, early access, doi.org/10.1016/j.dcan.2022.02.008.

[8] M. Muhammad and G. A. Safdar, "Survey on existing authentication issues for cellular-assisted V2X communication," *Vehicular Communications*, vol. 12, pp. 50–65, Feb. 2018.

[9] B. Li, Z. Fei, C. Zhou and Y. Zhang, "Physical-layer security in space information networks: A survey," *IEEE Internet of Things Journal*, vol. 7, no. 1, pp. 33-52, Jan. 2020.

[10] B. He, F. Wang, Y. Liu and S. Wang, "Specific emitter identification via multiple distorted receivers," in *IEEE International Conference on Communications Workshops (ICC Workshops)*, 2019, pp. 1–6.

[11] B. He and F. Wang, "Cooperative specific emitter identification via multiple distorted receivers," *IEEE Transactions on Information Forensics and Security*, vol. 15, pp. 3791–3806, Jun. 2020.

[12] Q. Li, X. Sun, P. Lv and C. Sun, "The identification of power IoT devices on differential constellation trajectory map," in *International Conference on Measuring Technology and Mechatronics Automation (ICMTMA)*, 2021, pp. 140–144.

[13] Y. Sun, S. Abeywickrama, L. Jayasinghe, C. Yuen, J. Chen and M. Zhang, "Micro doppler signature based detection, classification, and localization of small UAV with long short term memory neural network," *IEEE Trans. on Geoscience and Remote Sensing*, vol. 59, Aug. 2021, pp. 6285–6300.

[14] H. Ye, G. Y. Li and B.-H. Juang, "Power of deep learning for channel estimation and signal detection in OFDM systems," *IEEE Wireless Communications Letters*, vol. 7, no. 1, pp. 114–117, Feb. 2018.

[15] H. He, C. Wen, S. Jin and G. Y. Li, "Deep learning-based channel estimation for beamspace mmWave massive MIMO systems," *IEEE Wireless Communications Letters*, vol. 7, no. 5, pp. 852–855, Oct. 2018.

[16] W. Xia, G. Zheng, Y. Zhu, J. Zhang, J. Wang and A. P. Petropulu, "A deep learning framework for optimization of MISO downlink beamforming," *IEEE Transactions on Communications*, vol. 68, no. 3, pp. 1866–1880, Mar. 2020.

[17] C. Wen, W. Shih and S. Jin, "Deep learning for massive MIMO CSI feedback," *IEEE Wireless Communications Letters*, vol. 7, no. 5, pp. 748–751, Oct. 2018.

[18] L. Liang, H. Ye, G. Yu and G. Y. Li, "Deep-learning-based wireless resource allocation with application to vehicular networks," *Proceedings of the IEEE*, vol. 108, no. 2, pp. 341-356, Feb. 2020.

[19] K. Merchant, S. Revay, G. Stantchev and B. Nousain, "Deep learning for RF device fingerprinting in cognitive communication networks," *IEEE Journal of Selected Topics in Signal Processing*, vol. 12, no. 1, pp. 160–167, Feb. 2018.

[20] J. Yu, A. Hu, G. Li and L. Peng, "A robust RF fingerprinting approach using multisampling convolutional neural network," *IEEE Internet of Things Journal*, vol. 6, no. 4, pp. 6786–6799, Aug. 2019.

[21] X. Wang, Y. Zhang, H. Zhang, Y. Li and X. Wei, "Radio frequency signal identification using transfer learning based on LSTM," *Circuits, Systems, and Signal Processing*, vol. 39, no.11, pp. 5514–5528, Apr. 2020.

[22] B. He and F. Wang, "Cooperative specific emitter identification via multiple distorted receivers," *IEEE Transactions on Information Forensics and Security*, vol. 15, pp. 3791–3806, June 2020.

[23] T. Jian, Y. Gong, *et al*., “Radio frequency fingerprinting on the edge,” *IEEE Transactions on Mobile Computing*, to be published, doi: 10.1109/TMC.2021.3064466.

[24] Y. Wang, G. Gui, H. Gacanin, T. Ohtsuki, O. A. Dobre and H. V. Poor, “An efficient specific emitter identification method based on complex-valued neural networks and network compression,” *IEEE Journal on Selected Areas in Communications*, vol. 39, no. 8, pp. 2305–2317, Aug. 2021.

[25] L. Ding, S. Wang, F. Wang and W. Zhang, “Specific emitter identification via convolutional neural networks,” *IEEE Communications Letters*, vol. 22, no. 12, pp. 2591–2594, Dec. 2018.

[26] G. Shen, J. Zhang, *et al*., “Radio frequency fingerprint identification for LoRa using spectrogram and CNN,” in *IEEE Conference on Computer Communications*, 2021, pp. 1–10.

[27] C. Xie, L. Zhang, Z. Zhong, “Virtual adversarial training-based semisupervised specific emitter identification,” *Wireless Communications and Mobile Computing*, in press, doi.org/10.1155/2022/6309958.

[28] J. Gong, X. Xu, and Y. Lei, “Unsupervised specific emitter identification method using radio-frequency fingerprint embedded InfoGAN,” *IEEE Transactions on Information Forensics and Security*, vol. 15, pp. 2898–2913, Mar. 2020.

[29] L. Peng, J. Zhang, M. Liu and A. Hu, “Deep learning based RF fingerprint identification using differential constellation trace figure,” *IEEE Transactions on Vehicular Technology*, vol. 69, no. 1, pp. 1091–1095, Jan. 2020.

[30] P. Yin, L. Peng, *et al*.,, “LTE device identification based on RF fingerprint with multi-channel convolutional neural network,” in *IEEE Global Communications Conference*, 2021, pp. 1–6.

[31] Y. Peng, P. Liu, Y. Wang, *et al*., “Radio frequency fingerprint identification based on slice integration cooperation and heat constellation trace figure,” *IEEE Wireless Communications Letters*, vol. 11, no. 3, pp. 543–547, Mar. 2022.

[32] H. Fu, S. Abeywickrama, L. Zhang and C. Yuen, “Low complexity portable passive drone surveillance via SDR-based signal processing,” *IEEE Communications Magazine*, vol. 56, Apr. 2017, pp. 112–118.

[33] A. Aghnaiya, Y. Dalveren and A. Kara, “On the performance of variational mode decomposition-based radio frequency fingerprinting of Bluetooth devices,” *Sensors*, vol.20, no. 6, pp. 1704–1713, Mar. 2020.

[34] Y. Huang, “Radio frequency fingerprint extraction of radio emitter based on I/Q imbalance,” *Procedia Computer Science*, vol. 107, pp. 472–477, Apr. 2017.

[35] B. Danev, T. S. Heydt-Benjamin and S. Capkun, “Physical-layer identification of RFID devices,” in *USENIX Security Symposium*, 2009, pp. 199–214.

[36] Y. Huang, and H. Zheng, “Radio frequency fingerprinting based on the constellation errors,” in *IEEE Asia-Pacific Conference on Communications (APCC)*, 2012, pp. 900–905.

[37] L. Li, H. Ji and L. Wang, “Specific radar emitter recognition based on wavelet packet transform and probabilistic SVM,” in *IEEE International Conference on Information and Automation*, 2009, pp. 1308–1313.

[38] Y. Yuan, Z. Huang, H. Wu and X. Wang “Specific emitter identification based on Hilbert-Huang transform-based time-frequency-energy distribution features,” *IET communications*, vol. 8, no. 13, pp. 2404–2412, Apr. 2014.

[39] Y. Tu, Z. Zhang, Y. Li, C. Wang and Y. Xiao, “Research on the Internet of things device recognition based on RF-fingerprinting,” *IEEE Access*, vol. 7, pp. 37426–37431, Mar. 2019.

[40] J. Lu, P. Gong, J. Ye and C. Zhang, “Learning from very few samples: A survey,” [online] available: https://arxiv.org/abs/2009.02653.

[41] H. Zhou, *et al*., “Few-shot electromagnetic signal classification: A data union augmentation method,” *Chinese Journal of Aeronautics*, early access, doi.org/10.1016/j.cja.2021.07.014.

[42] Y. Wang, L. Yao, Y. Wang and Y. Zhang, "Robust CSI-based human activity recognition with augment few shot learning," *IEEE Sensors Journal*, vol. 21, no. 21, pp. 24297–24308, Nov. 2021.

[43] L. Bertinetto, J. F. Henriques, A. Vedaldi, and P. Torr, "Fully-convolutional siamese networks for object tracking," *European conference on computer vision*, 2016, pp. 850–865.

[44] Y. Wu, Z. Sun and G. Yue, "Siamese network-based open set identification of communications emitters with comprehensive features," in *IEEE International Conference on Communication, Image and Signal Processing (CCISP)*, 2021, pp. 408-412.

[45] G. Sun, "RF transmitter identification using combined Siamese networks," *IEEE Transactions on Instrumentation and Measurement*, early access, doi: 10.1109/TIM.2021.3135005.

[46] Z. Zhang, Y. Li and M. Gao, "Few-shot learning of signal modulation recognition based on attention relation network," *European Signal Processing Conference (EUSIPCO)*, 2021, pp. 1372–1376.

[47] P. Man, C. Ding, W. Ren and G. Xu, "A specific emitter identification algorithm under zero sample condition based on metric learning," *Remote Sensing*, vol. 13, no. 23, pp. 4919–4928, Dec. 2021.

[48] N. Yang, B. Zhang, G. Ding, *et al*., "Specific emitter identification with limited samples: A model-agnostic meta-learning approach," *IEEE Communications Letters*, early access, doi: 10.1109/LCOMM.2021.3110775.

[49] Y. Dong, Y. Peng, M. Yang, S. Lu and Q. Shi, "Signal transformer: Complex-valued attention and meta-learning for signal recognition," [online] avilable: https://arxiv.org/abs/2106.04392, 2021.

[50] A. Antoniou, H. Edwards and A. Storkey, "How to train your MAML," [online] avilable: https://arxiv.org/abs/1810.09502, 2018.

[51] Y. Tu, Y. Lin, *et al*., "Large-scale real-world radio signal recognition with deep learning," *Chinese Journal of Aeronautics*, early access, doi: 10.1016/j.cja.2021.08.016.

[52] E. Hoffer and N. Ailon, "Deep metric learning using triplet network," in *International workshop on similarity-based pattern recognition*, 2015, pp. 84–92.

[53] Y. Wen, *et al*., "A discriminative feature learning approach for deep face recognition," in *European conference on computer vision*, 2016, pp. 499–515.

[54] J. Liu, *et al*., "Multi-scale triplet CNN for person re-identification," in *ACM international conference on Multimedia*, 2016, pp. 192–196.

[55] X. He, Y. Zhou, Z. Zhou, S. Bai and X. Bai, "Triplet-center loss for multi-view 3D object retrieval," in *IEEE Conference on Computer Vision and Pattern Recognition*, 2018, pp. 1945–1954.

[56] S. Aranganayagi and K. Thangavel, "Clustering categorical data using silhouette coefficient as a relocating measure" in *International Conference on Computational Intelligence and Multimedia Applications*, 2007, pp. 13–17.

[57] T. Gao, Z. Liu and M. Sun, "Hybrid attention-based prototypical networks for noisy few-shot relation classification," in *AAAI Conference on Artificial Intelligence*, 2019, pp. 6407–6414.